\documentclass[twocolumn]{revtex4-1}
\usepackage{amsmath, amsfonts, graphicx, verbatim, hyperref, subfigure, float}

\begin{document}

\title{Quantum Particle Production Effects on Cosmic Expansion}

\author{Fernando Zago}
\email{fernando.zago@pitt.edu}

\author{Arthur Kosowsky}
\email{kosowsky@pitt.edu}

\affiliation{Department of Physics and Astronomy, University of Pittsburgh, Pittsburgh, PA 15260 USA and \\ Pittsburgh Particle Physics, Astrophysics, and Cosmology Center (PITT PACC), Pittsburgh PA 15260}

\begin{abstract}
Quantum fields in cosmological spacetimes can experience particle production due to their interaction with the expanding background. This effect is particularly relevant for models of the very early Universe, when the energy density generated through this process may back-react on the cosmological expansion. Yet, these scenarios have not been fully explored due to the several technical hurdles imposed by the back-reaction calculations. In this work we review the basics of cosmological quantum particle production and demonstrate a numerical algorithm to solve the back-reaction problem in regimes dominated by particle production. As an illustration, we compute the effects of a massive quantized scalar field on a cosmological bounce scenario, explicitly showing that quantum particle production can cause the contracting phase to end in a radiation crunch, or can delay the bounce. Finally, we discuss the relevance of quantum particle production/annihilation to bounce and inflationary models of the early Universe.
\end{abstract}


\maketitle

\section{Introduction}\label{Sec. Introduction}

In his pioneering 1968 Ph.D. thesis, Leonard Parker discovered the surprising phenomenon that evolving cosmological spacetimes can produce quantum particles \cite{Parker69, Parker71, ZeldovichStarobinsky77}. This work laid much of the theoretical framework for our current understanding of quantum fields in curved spacetimes (see, e.g., the textbooks \cite{ParkerToms09, BirrellDavies82}). The phenomenon has since been investigated in  astrophysical and cosmological contexts, leading to fundamental theoretical results including the emission of Hawking radiation by blackholes \cite{Hawking75, Unruh76, Davies76, Page82, Brown86, Frolov87, Anderson93, Anderson94, Anderson95} and the generation of primordial fluctuations during inflation \cite{Starobinsky79, Allen88, Sahni90, Mukhanov92, Souradeep92, Glenz09, Agullo11}.

Cosmological quantum particle production results from the shifting quantum vacuum state due to spacetime expansion. A given mode of a quantum field can begin in a state with no particles, but at a later time have a non-zero particle number expectation value. Parker showed that while this does not happen for massless conformally-coupled fields, it is generic for massive scalar fields of arbitrary coupling to the spacetime curvature. Typically, the particle production is significant when the particle mass $m$ is of the order of the expansion rate $H$, or when $m^{2} \simeq \dot{H}$. Simple dimensional arguments show that in a Universe with critical density $\rho_c \simeq 3H^{2}/8\pi G$, quantum particle production can contribute significantly to the energy density of the Universe at early epochs when $H$ is not too far below the Planck scale. 

This effect naturally suggests that quantum particle production may have a significant impact on the expansion rate of the Universe at early times. The back-reaction problem consists in understanding how the energy density generated through this process alters the evolution of the background spacetime. This seemingly straightforward calculation is actually subtle due to two technical challenges of quantum fields in curved spacetimes. First, the definition of the quantum vacuum and particle are not formally well defined in time-varying spacetimes. Second, the energy-momentum tensor of a quantized field propagating on a curved background possesses a formally infinite expectation value, and must be regularized to yield physically sensible results. In homogeneous and isotropic spacetimes, both problems can be addressed via the method of adiabatic regularization \cite{ParkerFulling74, FullingParker74, Bunch80, Fulling74, Christensen78, AndersonParker87}, which is particularly useful for numerical computations. The adiabatic notion of particle offers a clear way to track particle number in a spacetime which is expanding sufficiently slowly. This representation has the special property of defining a vacuum state which comes closest to the Minkowski vacuum when the background expansion is sufficiently slow. However, adiabatic regularization introduces several technical complications to the back-reaction problem. As a consequence, ambiguities arise in specifying initial conditions, and computationally the problem becomes very complicated with potential numerical instabilities \cite{Anderson83, Anderson84, Anderson85, Anderson86, SuenAnderson87}. In practice, the ambiguities and complexities together have prevented any general numerical solution, although the test-field limit in which back-reaction effects are neglected has been investigated in several studies \cite{Birrell78, Anderson00, Anderson05, Bates10, Habib99}.

Formal developments clarified aspects of the adiabatic regularization approach, offering a clear separation between the energy density due to the field particle content and the divergent contributions from the zero-point energy \cite{Habib99, AndersonMottola14, AndersonMottola14b, Anderson18}. However, the notion of adiabatic particle appearing in the energy density suffers from ambiguities. For a typical mode of a quantum field, its associated particle number during times of significant particle production depends on the perturbative order of adiabatic regularization employed.  This is obviously an unphysical result, since the dominant term in the energy density is often just a simple function of the particle number density. Typically, successive orders of adiabatic regularization gives the particle number in a given mode as a divergent asymptotic series. Recent important papers by Dabrowski and Dunne \cite{DabrowskiDunne14, DabrowskiDunne16} employed a remarkable result of asymptotic analysis \cite{Dingle73, Berry90, Berry82, Berry89, Berry88} to sum the divergent series and provide a sensible notion of particle which is valid at all times. Since any remaining physical ambiguity is then removed from the problem, this result points the way to a general numerical solution.

We combine these recent results into an approach which numerically solves the quantum back-reaction problem in regimes with field energy density dominated by particle production. We then apply this technique to a toy cosmological model, that of a positive-curvature spacetime with a constant energy density (the closed de Sitter model). In the absence of any quantum fields or other particle content, this spacetime exhibits a bounce behavior, contracting to a minimum scale factor and then expanding again. Here we show explicitly that the existence of a massive scalar field in a particular mass range will cause large changes in the spacetime evolution: even if the contracting spacetime initially contains a quantum field in its adiabatic vacuum state, quantum particle production can create enough energy density to push the spacetime into a radiation crunch. Special values of the field mass can also delay but not eliminate the bounce. This appears to be the first general solution for the quantum back-reaction problem in cosmology.

In Section II we review standard results for quantized scalar fields propagating in cosmological spacetimes, while Section III discusses the adiabatic field representation and the semi-classical notion of adiabatic particle number. Section IV recasts quantum particle production in terms of the Stokes phenomenon of the complex-plane wave equation for specific modes, including interference between different modes. Section V formulates the back-reaction problem for scenarios in which the field particle content or particle production dominates the field energy density. Section VI outlines our numerical implementation of the mathematical results in Sections IV and V. Physical results for a closed de Sitter model are presented in Section VII. Finally, in Section VIII we discuss the prospects for more general situations, including quantum fields with spin and interacting quantum fields, and the possible relevance of quantum particle production to early-Universe models, including inflationary and bounce scenarios. Salient technical details are summarized in the Appendix. Natural units with $\hbar = c = 1$ are adopted throughout.

\section{Scalar Fields in FLRW Spacetimes}\label{Sec. QFT in FLRW}

We first summarize basic results for scalar fields in spatially isotropic and homogeneous spacetimes (see, e.g., \cite{ParkerToms09, BirrellDavies82}). 
Consider a Universe described by the Friedmann-Lema\^{i}tre-Robertson-Walker (FLRW) metric
\begin{equation}\label{Eq. FLRW Metric}
	\mathrm{d}s^{2} = g_{ab}\mathrm{d}x^{a}\mathrm{d}x^{b} = -\mathrm{d}t^{2} + a^{2}(t) g_{ij}\mathrm{d}x^{i}\mathrm{d}x^{j} \, ,
\end{equation}
with
\begin{equation}\label{Eq. FLRW Spatial Metric}
	g_{ij}\mathrm{d}x^{i}\mathrm{d}x^{j} =\frac{\mathrm{d}r^{2}}{1 - K r^{2}} + r^{2} \mathrm{d}\theta^{2} + r^{2} \sin^{2}\theta \, \mathrm{d}\varphi^{2} \, .
\end{equation}
Here $a(t)$ is the scale factor which describes the cosmological expansion history, and $K = -1, \, 0, \, +1$ corresponds to the curvature parameter of an open, flat, and closed Universe, respectively.
The non-vanishing components of the Ricci tensor $R_{ab}$ are
\begin{subequations}\label{Eq. Ricci Tensor}
\begin{align}
	R_{00}(t) &= 3\Big[ \dot{H}(t) + H^{2}(t) \Big] g_{00} \, ,\\[8pt]
	R_{ij}(t) &= \Big[ \dot{H}(t) + 3 H^{2}(t) + 2K/a^{2}(t) \Big] g_{ij} \, ,
\end{align}
\end{subequations}
and the Ricci scalar $R = g^{ab}R_{ab}$ is
\begin{equation}\label{Eq. Ricci Scalar}
	R(t) = 6 \Big[ \dot{H}(t) + 2 H^{2}(t) + K/a^{2}(t) \Big] \, 
\end{equation}
where
\begin{equation}\label{Eq. Hubble Parameter}
	H(t) \equiv \frac{\dot{a}(t)}{a(t)} 
\end{equation}
is the Hubble parameter. Overdots indicate differentiation with respect to proper time $t$.

We are interested in the evolution of a free scalar field $\Phi(t,\, \mathbf{x})$ of arbitrary mass and curvature coupling in this spacetime. The action for such a field can be expressed generically as
\begin{align}\label{Eq. Field Action}
	S = -\frac{1}{2} \int \sqrt{-g} \, \mathrm{d}^{4}x \Big[ \big( \nabla_{a} \Phi \big) g^{ab} \big( \nabla_{b} \Phi \big) + \Big. & \\
	 \Big. + m^{2} \Phi^{2} + \xi R \Phi^{2} \Big] & \, , \nonumber
\end{align}
where $\nabla_{a}$ is the covariant derivative, $g = \det{(g_{ab})}$, $m$ is the field mass, and $\xi$ is the field coupling to the spacetime curvature. Applying the variational principle to this action yields the equation of motion 
\begin{equation}\label{Eq. Field Equation of Motion}
	\Big[ \Box - m^{2} - \xi R(t) \Big]\Phi(t,\, \mathbf{x}) = 0 \, ,
\end{equation}
where $\Box = g^{ab}\nabla_{a}\nabla_{b}$ is the d'Alembert operator associated with the spacetime.

Due to the homogeneity and isotropy of the background metric, the solutions of Eq.~(\ref{Eq. Field Equation of Motion}) 
can be separated into purely temporal and spatial parts. As a consequence, the quantized field operator can be written as
\begin{align}\label{Eq. Field}
	\hat{\Phi}(t,\, \mathbf{x})=a^{-3/2}(t) \! \int \mathrm{d}\mu(k) \Big[ a_{\mathbf{k}}^{\phantom{\dagger}}f_{k}(t)Y_{\mathbf{k}}(\mathbf{x}) + \Big. & \\
	\Big. + a_{\mathbf{k}}^{\dagger}f_{k}^{\ast}(t)Y_{\mathbf{k}}^{\ast}(\mathbf{x}) \Big] & \, , \nonumber
\end{align}
where the raising and lowering operators $a_{\mathbf{k}}^{\dagger}$ and $a_{\mathbf{k}}^{\phantom{\dagger}}$ satisfy the canonical commutation relations
\begin{equation}\label{Eq. Commutation Relations}
\left[ a_{\mathbf{k}}^{\phantom{\dagger}} \, , a_{\mathbf{k^{\prime}}}^{\dagger} \right] = \delta_{\mathbf{k},\, \mathbf{k^{\prime}}} \, ,
\end{equation}
and $\mathrm{d}\mu(k)$ is a geometry-dependent integration measure given by 
\begin{equation}\label{Eq. Integration Measure}
	\int \mathrm{d}\mu(k) =
	\left\{\begin{aligned}
			&\sum_{k=1}^{\infty} k^2\, , && \text{for } K = +1 \\
			&\int_{0}^{\infty} k^2 \,\mathrm{d}k \, , && \text{for } K = 0,\,-1 \, .
		\end{aligned}\right.
\end{equation}
The functions $Y_\mathbf{k}(\mathbf{x})$ and $f_k(t)$ contain the spatial and temporal dependence of each $\mathbf{k}$-mode. The harmonic functions $Y_\mathbf{k}(\mathbf{x})$ are eigenfunctions of the Laplace-Beltrami operator associated with the geometry of spatial hypersurfaces, while the mode functions $f_k(t)$ obey the harmnonic oscillator equation
\begin{equation}\label{Eq. Mode Equation}
\ddot{f}_{k}(t) + \Omega_{k}^{2}(t) f_{k}(t) = 0 \, 
\end{equation}
with the time-dependent frequency function 
\begin{equation}\label{Eq. Mode Frequency}
\Omega_{k}^{2}(t) = \omega_{k}^{2}(t) + \bigg( \xi - \frac{1}{6} \bigg) R(t) - \Bigg[ \frac{\dot{H}(t)}{2} + \frac{H^{2}(t)}{4} \Bigg] \, ,
\end{equation}
where
\begin{equation}\label{Eq. Minkowski Frequency}
	 \omega_{k}(t) = \Bigg[\frac{k^{2}}{a^{2}(t)} + m^{2}\Bigg]^{\!1/2} \, .
\end{equation}
The complex mode functions $f_{k}(t)$ and $f^{\ast}_{k}(t)$ also satisfy the Wronskian condition
\begin{equation}\label{Eq. Wronskian Condition}
f_{k}(t)\dot{f}^{\ast}_{k}(t) - \dot{f}_{k}(t) f^{\ast}_{k}(t) = i \, .
\end{equation}
If Eq.~(\ref{Eq. Wronskian Condition}) holds at some particular time $t$, then Eq.~(\ref{Eq. Mode Equation}) guarantees it will also hold at all future
times.

The quantization procedure outlined above naturally leads to the construction of the Fock space of field states. The base element of this space is the vacuum state, which is defined as the normalized state that is annihilated by all lowering operators:
\begin{equation}\label{Eq. Vacuum}
a_{\mathbf{k}}^{\phantom{\dagger}} \left| 0 \right\rangle = 0 \ \ \mathrm{and} \ \ \left\langle 0 | 0 \right\rangle = 1 \, .
\end{equation}
All remaining states are generated from the vacuum by the successive application of raising operators, such as
\begin{equation}
\left| \mathbf{k}_1 ,\, \mathbf{k}_2 ,\, \dots \right\rangle = a_{\mathbf{k}_1}^{\dagger} a_{\mathbf{k}_2}^{\dagger} \dots \left| 0 \right\rangle \, ,
\end{equation}
and normalized by the requirement of mutual orthonormality. In what follows, we will be interested in the family of field states which are spatially isotropic and homogeneous, as these constitute viable sources of the FLRW metric.

The field operator $\hat{\Phi}(t,\, \mathbf{x})$ admits numerous representations of the form shown in Eq.~(\ref{Eq. Field}), each of which is associated with a different mode function pertaining to the set of solutions of Eq.~(\ref{Eq. Mode Equation}). These representations are related: the complex mode functions $f_{k}(t)$ and $h_{k}(t)$ belonging to any two different representations can be expressed in terms of one another through the Bogolyubov transformations
\begin{subequations}\label{Eq. Bogolyubov Modes}
\begin{align}
	{f}_{k}(t) &= \alpha_{k}h_{k}(t) + \beta_{k}h^{\ast}_{k}(t) \, ,\\[8pt]
	{f}_{k}^{\ast}(t) &= \beta_{k}^{\ast}h_{k}(t) + \alpha_{k}^{\ast}h^{\ast}_{k}(t) \, ,
\end{align}
\end{subequations}
where $\alpha_{k}$ and $\beta_{k}$ are known as Bogolyubov coefficients. Due to homogeneity and isotropy, these coefficients depend only on ${k = \left| \mathbf{k} \right|}$. Substituting these expressions into Eq.~(\ref{Eq. Field}) leads to similar transformations relating the raising and lowering operators belonging to these representations:
\begin{subequations}\label{Eq. Bogolyubov Operators}
\begin{align}
a_{\mathbf{k}}^{\phantom{\dagger}} &= \alpha_{k}^{\ast} b_{\mathbf{k}}^{\phantom{\dagger}} - \beta_{k}^{\ast} b_{\mathbf{k}}^\dagger\, , \\[8pt]
a_{\mathbf{k}}^\dagger &= \alpha_{k} b_{\mathbf{k}}^\dagger - \beta_{k} b_{\mathbf{k}}^{\phantom{\dagger}} \, ,
\end{align}
\end{subequations}
from which it follows that the Bogolyubov coefficients must satisfy
\begin{equation}\label{Eq. Bogolyubov Constraint}
	\big| \alpha_{k} \big|^{2} - \big| \beta_{k} \big|^{2} = 1 \,
\end{equation}
in order to guarantee that the commutation relations of Eq.~(\ref{Eq. Commutation Relations}) are valid across all representations.

A direct consequence of Eqs.~(\ref{Eq. Bogolyubov Operators}) is that the notion of vacuum is not unique for a quantized field defined on a FLRW spacetime \cite{BirrellDavies82, ParkerToms09}. This is evident from the following simple calculation, which shows that the vacuum defined in Eq.~(\ref{Eq. Vacuum}) is not necessarily devoid of particles according to the number operator belonging to a different field representation:
\begin{align}
	\mathcal{N}_{k} &= \big\langle 0 \big| b_{\mathbf{k}}^{\dagger} b_{\mathbf{k}}^{\phantom{\dagger}} \big| 0 \big\rangle \nonumber \\
	&= \big| \alpha_{k} \big|^{2} \big\langle 0 \big| a_{\mathbf{k}}^{\dagger} a_{\mathbf{k}}^{\phantom{\dagger}}  \big| 0 \big\rangle + \big| \beta_{k} \big|^{2} \big\langle 0 \big| a_{\mathbf{k}}^{\phantom{\dagger}} a_{\mathbf{k}}^\dagger \big| 0 \big\rangle \\
	&= \big| \beta_{k} \big|^{2} \nonumber \, .
\end{align}
Therefore, different choices of representation inevitably lead to distinct notions of vacuum and, consequently, to distinct notions of particle. This result is a quite general feature of quantum field theory defined on curved spacetimes, and although it initially seems troublesome, it actually becomes useful in numerical back-reaction calculations. To that end, we introduce in the next section a particularly useful representation which defines the most physical notion of particle in a FLRW spacetime.

\section{Adiabatic Representation}\label{Sec. Adiabatic Representation}

Despite the multitude of available representations for a quantized scalar field defined on a FLRW spacetime, one particular choice referred to as the adiabatic representation stands out. This representation has the special property of defining a vacuum state which comes closest to the Minkowski vacuum when the background expansion is sufficiently slow. As a consequence, the adiabatic representation provides the most meaningful notion of physical particle in an expanding homogeneous and isotropic Universe. Here we discuss this representation 
closely following Ref.~\cite{Habib99}.

The adiabatic representation is characterized by mode functions which are the phase-integral solutions \cite{Parker69, Parker71, ZeldovichStarobinsky77} of Eq.~(\ref{Eq. Mode Equation}):
\begin{equation}\label{Eq. Adiabatic Modes}
	h_{k}(t) = \frac{1}{\sqrt{2 W_{k}(t)}} \exp{\!\bigg( \! -i \! \int^{t} W_{k}(s) \, \mathrm{d}s \bigg)} \, ,
\end{equation}
where the integral in the exponent can be computed from any convenient reference time, and the function $W_{k}(t)$ is given by the formal asymptotic series
\begin{equation}\label{Eq. Adiabatic W}
	 W_{k}(t) \equiv \Omega_{k}(t) \sum_{n=0}^{\infty} \varphi_{k,\,2n}(t) \, .
\end{equation}
The terms $\varphi_{k,\,2n}(t)$ are obtained by substituting Eqs.~(\ref{Eq. Adiabatic W}) and (\ref{Eq. Adiabatic Modes}) into Eq.~(\ref{Eq. Mode Equation}). The expressions which ensue from these substitutions are standard results of the phase-integral method \cite{FromanFroman96, FromanFroman02}; up to fourth order they are
\begin{subequations}
\begin{align}
	&\varphi_{k,\,0}(t) = 1 \, , \\[8pt]
	&\varphi_{k,\,2}(t) = \frac{1}{2} \varepsilon_{k,\,0}(t) \, , \\[8pt]
	&\varphi_{k,\,4}(t) = -\frac{1}{8}\Big[ \varepsilon_{k,\,0}^{2}(t) + \varepsilon_{k,\,2}(t) \Big] \, ,
\end{align}
\end{subequations}
for which the quantities appearing on the right-hand sides are given by
\begin{subequations}
\begin{align}
	\varepsilon_{k,\,0}(t) &\equiv \Omega_{k}^{-3/2}(t) \, \frac{d^{2}}{dt^{2}} \! \bigg[ \Omega_{k}^{-1/2}(t) \bigg]\, , \\[8pt]
	\varepsilon_{k,\,m}(t) &\equiv \bigg[ \Omega_{k}^{-1}(t) \, \frac{d}{dt} \bigg]^{\!m} \!\!\varepsilon_{k,\,0}(t) \, .
\end{align}
\end{subequations}

In a sense, the functions $W_{k}(t)$ capture the overall time dependence of each $\mathbf{k}$-mode due the evolving FLRW metric, leaving behind only the Minkowski-like mode oscillations which take place on top of this background \cite{DabrowskiDunne14, DabrowskiDunne16}. It is this property that makes the adiabatic mode functions $h_{k}(t)$ and $h^{\ast}_{k}(t)$ such good templates for probing the particle content of fields evolving in cosmological spacetimes. This template role is made precise by the following time-dependent generalization of Eqs.~(\ref{Eq. Bogolyubov Modes}) \cite{Parker69, Parker71, ZeldovichStarobinsky77}, which expresses the field modes $f_{k}(t)$ as linear combinations of the adiabatic mode functions:
\begin{equation}\label{Eq. Adiabatic Bogolyubov Mode}
	{f}_{k}(t) = \alpha_{k}(t)h_{k}(t) + \beta_{k}(t)h^{\ast}_{k}(t) \, ,
\end{equation}
where the Bogolyubov coefficients $\alpha_{k}(t)$ and $\beta_{k}(t)$ are analogous to those appearing in Eqs.~(\ref{Eq. Bogolyubov Modes})~and~(\ref{Eq. Bogolyubov Operators}), but are here regarded as time-dependent quantities due to the fact that $h_{k}(t)$ and $h_{k}^{\ast}(t)$ are merely approximate solutions of Eq.~(\ref{Eq. Mode Equation}). In order to completely specify these coefficient functions, an additional expression must be provided. For that purpose, it is common to introduce a condition on the time derivative of the mode function which preserves the Wronskian relation of Eq.~(\ref{Eq. Wronskian Condition}). In its most general form, this condition can be stated as \cite{Habib99}
\begin{align}\label{Eq. Adiabatic Bogolyubov Mode Derivative}
	\dot{f}_{k}(t) = \bigg[ -iW_{k}(t) + \frac{V_{k}(t)}{2}  \bigg] \alpha_{k}(t)h_{k}(t) \,\, & \\
	 + \, \bigg[ iW_{k}(t) + \frac{V_{k}(t)}{2}  \bigg] \beta_{k}(t)h^{\ast}_{k}(t)& \, . \nonumber
\end{align}
Here the arbitrary function $V_{k}(t)$  contains the residual freedom in the definition of the adiabatic vacuum. In this work we will choose this function to be
\begin{equation}\label{Eq. Adiabatic V}
	V_{k}(t) = -\frac{\dot{W}_{k}(t)}{W_{k}(t)} \, ,
\end{equation}
as this choice leads to important simplifications in the back-reaction problem. 

Gathering Eqs.~(\ref{Eq. Field}) and (\ref{Eq. Adiabatic Bogolyubov Mode}), we find that the ladder operators associated with the adiabatic representation satisfy the transformations
\begin{subequations}\label{Eq. Bogolyubov Adiabatic Operators}
\begin{align}
a_{\mathbf{k}}^{\phantom{\dagger}} &= \alpha_{k}^{\ast}(t) b_{\mathbf{k}}^{\phantom{\dagger}}(t) - \beta_{k}^{\ast}(t) b_{\mathbf{k}}^{\dagger}(t)\, , \\[8pt]
a_{\mathbf{k}}^\dagger &= \alpha_{k}(t) b_{\mathbf{k}}^{\dagger}(t) - \beta_{k}(t) b_{\mathbf{k}}^{\phantom{\dagger}}(t) \, ,
\end{align}
\end{subequations}
which in turn imply a time-dependent version of Eq.~(\ref{Eq. Bogolyubov Constraint}), 
\begin{equation}\label{Eq. Adiabatic Bogolyubov Constraint}
	\big| \alpha_{k}(t) \big|^{2} - \big| \beta_{k}(t) \big|^{2} = 1 \, .
\end{equation}

Finally, it is useful to characterize field states according to the values of the non-trivial adiabatic bilinears $\big\langle b_{\mathbf{k}}^{\dagger}(t) b_{\mathbf{k}}(t) \big\rangle$ and $\big\langle b_{\mathbf{k}}(t) b_{\mathbf{k}}(t) \big\rangle$. The first of these bilinears tracks the adiabatic particle content per comoving volume in the $\mathbf{k}$-mode under consideration. Using the transformations established above by Eqs.~(\ref{Eq. Bogolyubov Adiabatic Operators}), it follows that
\begin{align}\label{Eq. Adiabatic Particles}
	\mathcal{N}_{k}(t) &= \big\langle b_{\mathbf{k}}^{\dagger}(t) b_{\mathbf{k}}(t) \big\rangle \nonumber \\
	&= \big| \alpha_{k}(t) \big|^{2} \big\langle a_{\mathbf{k}}^{\dagger} a_{\mathbf{k}}^{\phantom{\dagger}} \big\rangle + \big| \beta_{k}(t) \big|^{2} \big\langle a_{\mathbf{k}}^{\phantom{\dagger}} a_{\mathbf{k}}^\dagger \big\rangle \\
	&= N_{k} + \sigma_{k} \, \big| \beta_{k}(t) \big|^{2} \nonumber \, ,
\end{align}
where $N_{k} = \big\langle a_{\mathbf{k}}^{\dagger} a_{\mathbf{k}}^{\phantom{\dagger}} \big\rangle$ is a constant of motion which can be understood as the initial number of adiabatic particles per comoving volume populating the field mode of wavenumber $k$, and $\sigma_{k} = 1 + 2N_{k}$ is the Bose-Einstein parameter responsible for stimulated particle production. The second bilinear can be expressed as
\begin{align}\label{Eq. Adiabatic Interference}
	\mathcal{M}_{k}(t) &= \big\langle b_{\mathbf{k}}(t) b_{\mathbf{k}}(t) \big\rangle \nonumber \\
	&= \alpha_{k}(t) \beta^{\ast}_{k}(t) \big\langle a_{\mathbf{k}}^{\dagger} a_{\mathbf{k}}^{\phantom{\dagger}} \big\rangle + \alpha_{k}(t) \beta^{\ast}_{k}(t) \big\langle a_{\mathbf{k}}^{\phantom{\dagger}} a_{\mathbf{k}}^\dagger \big\rangle \\
	&= \sigma_{k} \, \alpha_{k}(t) \beta^{\ast}_{k}(t) \nonumber \, .
\end{align}

In principle, these bilinears contain all the required information to track the field evolution and, consequently, the time dependence of the field energy density and pressure. In practice, however, these quantities suffer from an irreducible ambiguity which is particularly pronounced when $\mathcal{N}_{k}(t)$ and $\mathcal{M}_{k}(t)$ incur rapid changes, such as when particle production occurs. 
The root of this issue can be traced back to the asymptotic representation of $W_{k}(t)$, which is usually handled by simply truncating the series in Eq.~(\ref{Eq. Adiabatic W}) at a finite order. 
However, the values of the bilinears depend strongly on where the series is truncated if they are rapidly changing (see Ref.~\cite{DabrowskiDunne14} for striking graphical representations).
In the next Section, we discuss a technique for finding the exact universal evolution for both adiabatic bilinears which the asymptotic series represents.

\section{Particle Production and the Stokes Phenomenon}\label{Sec. Stokes}

The adiabatic representation introduced in the previous section provides an accurate description of the bilinears $\mathcal{N}_{k}(t)$ and $\mathcal{M}_{k}(t)$ whenever $\left| \varepsilon_{k,\,0} \right| \ll 1$. The more severely this condition is violated, the more unreliable these adiabatic quantities become. Adiabatic particle production, for instance, coincides with the momentary violation of this condition, implying that the notion of particle remains uncertain until particle production ceases. Nonetheless, a universal notion of particle can be restored for all times when particle production events are understood in terms of the Stokes phenomenon.

\begin{figure}[t!]
    \includegraphics[width=0.49\textwidth]{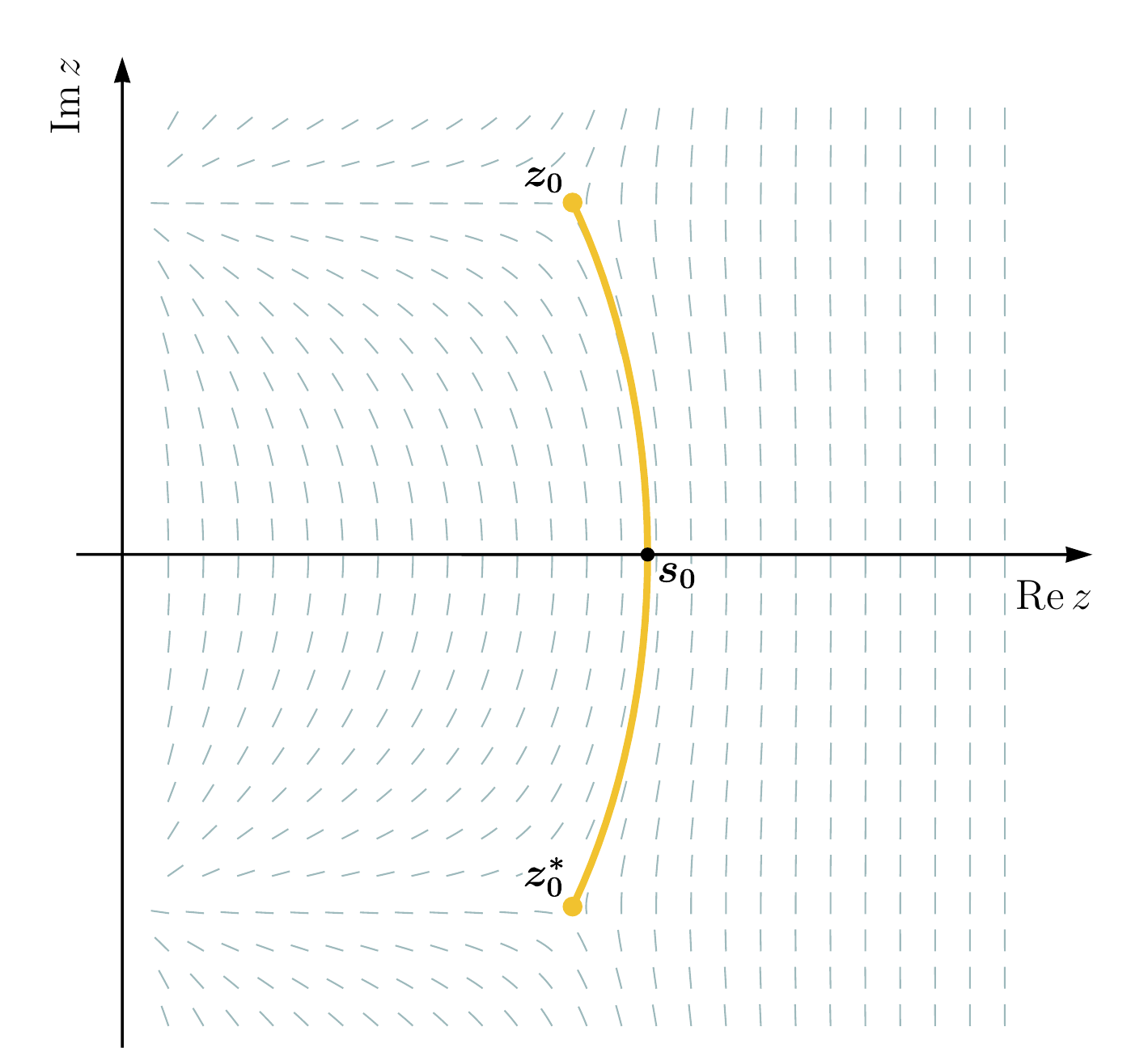}
    \caption{A depiction of the Stokes line sourced by a conjugate pair of simple turning points $\big( z_{0},\, z_{0}^{\ast}\big)$ of a frequency function $\Omega_{k}(z)$. Here the Stokes line crosses the real axis at the point $s_{0}$, which corresponds to the time at which particle production occurs for the mode of wavenumber $k$. The guiding lines show the directions for which the condition $\mathrm{Re}\big[\Omega_{k}\mathrm{d}z\big] = 0$ is locally satisfied.}
    \label{Fig. Stokes Line}
\end{figure}

The sharp transitions between asymptotic solutions of a given differential equation which are valid in different regions of the complex plane are termed the Stokes phenomenon. These regions are bounded by the so-called Stokes and anti-Stokes lines. In the context of a scalar field evolving in a FLRW spacetime, the differential equation of interest is the equation of motion for a given field mode extended to a complex time variable $z$:
\begin{equation}\label{Eq. Complex Mode Equation}
{f}^{\prime\prime}_{k}(z) + \Omega_{k}^{2}(z) f_{k}(z) = 0 \, ,
\end{equation}
in which the primes stand for differentiation with respect to $z$, the proper time is given by $t \equiv \mathrm{Re}\, z$, and $\Omega_{k}(z)$ represents the analytic continuation of the time-dependent frequency of Eq.~(\ref{Eq. Mode Frequency}). The Stokes lines associated with Eq.~(\ref{Eq. Complex Mode Equation}) are those lines which emanate from the zeros (also known as turning points) and poles of $\Omega_{k}(z)$ and along which $\mathrm{Re}\big[\Omega_{k}\mathrm{d}z\big]~\!\!=~\!\!0$. An illustration of such a line is shown in Fig.~\ref{Fig. Stokes Line}. The asymptotic solutions susceptible to the Stokes phenomenon are given by
\begin{equation}\label{Eq. Phase-Integral Solution}
	f_{k}(z) = \alpha_{k}(z) h_{k}(z) + \beta_{k}(z) h^{\ast}_{k}(z) \, ,
\end{equation}
where $h_{k}(z)$ and $h_{k}^{\ast}(z)$ are the complex extensions of the adiabatic mode functions defined in the previous section. As this solution evolves across a Stokes line, the values of the Bogolyubov coefficients $\alpha_{k}(z)$ and $\beta_{k}(z)$ change abruptly. By Eqs.~(\ref{Eq. Adiabatic Particles}) and (\ref{Eq. Adiabatic Interference}), this implies a sudden change in the adiabatic bilinears and, in particular, the production of adiabatic particles. Remarkably, a result from asymptotic analysis guarantees the existence of a smooth universal form for this rapid transition between different asymptotic regimes. Below we outline the derivation of this important result and summarize the quantities which determine the functional form of such smooth Stokes jumps.

We start by defining Dingle's singulant variable \cite{Dingle73} anchored at $z_{0}$:
\begin{equation}\label{Eq. Singulant}
F_{k}^{(0)}(z) = 2i \! \int_{z_0}^{z} \Omega_{k}(w) \, \mathrm{d}w \, ,
\end{equation}
where $z_0$ is a solution of $\Omega_{k}(z) = 0$ which sources the Stokes line of interest, is closest to the real axis, and is located in the upper half-plane. The singulant is a convenient variable for tracking the change incurred by the Bogolyubov coefficients $\alpha_{k}(z)$ and $\beta_{k}(z)$ across a Stokes line. Indeed, it was shown by Berry \cite{Berry90, Berry89, Berry88, Berry82} that these coefficients satisfy the following differential equations in the vicinity of a Stokes line:
\begin{subequations}\label{Eq. Bogolyubov Couplers}
\begin{align}
	\frac{d\beta_{k}}{dF_{k}^{(0)}} &= C_{\beta,\,k}^{(0)}\,\alpha_{k}\, , \\[8pt]
	\frac{d\alpha_{k}}{dF_{k}^{(0)}} &= C_{\alpha,\,k}^{(0)}\,\beta_{k}  \, ,
\end{align}
\end{subequations}
where $C_{\beta,\,k}^{(0)}$ and $C_{\alpha,\,k}^{(0)}$ are coupling functions which depend on the order at which the series representation of $W_{k}(z)$ is truncated. A remarkable discovery by Dingle \cite{Dingle73} states that the large-order terms in the asymptotic series of Eq.~(\ref{Eq. Adiabatic W}) have a closed form given by
\begin{equation}
\varphi_{k,\, 2n}(z) \sim -\frac{(2n - 1)!}{\pi F_{k}^{(0)\,2n}} \quad \text{for} \quad n \gg 1 \, .
\end{equation}
It is clear from this result that the smallest term in such a series corresponds to $n \approx \big| F_{k}^{(0)} \big|$. Terminating the series at this order leads to optimal closed form expressions for $C_{\beta,\,k}^{(0)}$ and $C_{\alpha,\,k}^{(0)}$, which can be substituted in Eqs.~(\ref{Eq. Bogolyubov Couplers}) to yield the following universal behaviors for $\beta_{k}(t)$ and $\alpha_{k}(t)$ along the real axis and across the Stokes line under consideration:
\begin{subequations}\label{Eq. Smooth Bogolyubov}
\begin{align}
	\beta_{k}(t) &\approx \frac{i}{2} \, \mathrm{Erfc}\Big( \! -\vartheta_{k}^{(0)}(t) \Big) \delta_{k}^{(0)}\, , \\[8pt]
	\alpha_{k}(t) &\approx \sqrt{1 + \big| \beta_{k}(t) \big|^{2}} \, ,
\end{align}
\end{subequations}
where $\vartheta_{k}^{(0)}(t)$ is a natural time evolution parameter which determines the sharpness of the Stokes jump, and $\delta_{k}^{(0)}$ corresponds to the jump's amplitude. Both of these parameters are expressible in terms of the singulant variable evaluated over the real axis:
\begin{subequations}
\begin{align}
	&\vartheta_{k}^{(0)}(t) = \frac{\phantom{\sqrt{2 \,\,}}\mathrm{Im} \Big[ F_{k}^{(0)}(t) \Big]}{\sqrt{2 \, \mathrm{Re} \Big[ F_{k}^{(0)}(t) \Big]}}\, , \\[8pt]
	&\delta_{k}^{(0)} = \exp\Big( \! -F_{k}^{(0)}(s_{0}) \Big) \, .
\end{align}
\end{subequations}
Here $F_{k}^{(0)}(s_{0})$ is simply the singulant computed at the point $z=s_{0}$, where the Stokes line sourced by $z_{0}$ intersects the real axis, i.e., 
\begin{equation}\label{Eq. Amplitude}
F_{k}^{(0)}(s_{0}) = 2i \! \int_{z_0}^{s_{0}} \Omega_{k}(w) \, \mathrm{d}w = i \! \int_{z_0}^{z_0^{\ast}} \Omega_{k}(w)\,\mathrm{d}w \, ,
\end{equation}
where the last equality follows from the reality of $\Omega_{k}(z)$ over the real axis. Putting together Eqs.~(\ref{Eq. Adiabatic Particles}), (\ref{Eq. Adiabatic Interference}), and (\ref{Eq. Smooth Bogolyubov}) yields a universal functional form which describes the time evolution of the adiabatic bilinears associated with the field mode of wavenumber $k$:
\begin{subequations}\label{Eq. Smooth Bilinears}
\begin{align}
	\mathcal{N}_{k} &\approx N_{k} + \frac{\sigma_{k}}{4} \Big| \mathrm{Erfc}\Big( \! -\vartheta_{k}^{(0)} \Big) \delta_{k}^{(0)} \Big|^{2}\, , \\[8pt]
	\mathcal{M}_{k} &\approx -i \, \frac{\sigma_{k}}{2} \Big[ \mathrm{Erfc}\Big( \! -\vartheta_{k}^{(0)} \Big) \delta_{k}^{(0)} \Big] \Big[ 1 + \mathcal{N}_{k} \Big]^{\!1/2}
\, .
\end{align}
\end{subequations}

These results can be further generalized to account for multiple Stokes line crossings, as well as the interference effects between them \cite{DabrowskiDunne14}. Define the accumulated phase between the first and the $p$-th pair of zeros of $\Omega_{k}(z)$ as
\begin{equation}\label{Eq. Phase}
\theta_{k}^{(p)} = \int_{s_0}^{s_p} \Omega_{k}(w)\,\mathrm{d}w 
\end{equation}
where $s_p$ corresponds to the point where the Stokes line associated with the $p$-th conjugate pair of zeros crosses the real axis. The functions which describe both adiabatic bilinears are then given by
\begin{subequations}\label{Eq. Bilinears Multiple Crossings}
\begin{align}
	\mathcal{N}_{k} &\approx N_{k} + \frac{\sigma_{k}}{4} \bigg| \sum_{p} \mathrm{Erfc}\Big( \! -\vartheta_{k}^{(p)} \Big) \delta_{k}^{(p)} \exp\Big(2i\theta_{k}^{(p)} \Big) \bigg|^{2}\, , \\[8pt]
	\mathcal{M}_{k} &\approx -i \, \frac{\sigma_{k}}{2} \bigg[ \sum_{p} \mathrm{Erfc}\Big( \! -\vartheta_{k}^{(p)} \Big) \delta_{k}^{(p)} \exp\Big( \! -2i\theta_{k}^{(p)} \Big) \bigg] \times \nonumber \\
	 & \,\,\;\; \quad \quad \times \! \bigg[ 1 + \mathcal{N}_{k} \bigg]^{\!1/2}
\end{align}
\end{subequations}
with $\delta_{k}^{(p)}$ and $\vartheta_{k}^{(p)}(t)$ being the amplitude and time evolution parameter associated with the $p$-th Stokes line.

Therefore, by monitoring the turning points and Stokes lines which accompany each mode's frequency function on the complex plane, we can track the evolution of the adiabatic bilinears related to any physically acceptable field state. In the next section we examine how this evolution affects the Universe's scale factor through the semi-classical Einstein equations.

\onecolumngrid
\section{The Semi-Classical Einstein Equations}\label{Sec. Semi-Classical Einstein}

If cosmological quantum particle production occurs at a sufficiently high rate, it can in principle back-react on the cosmic evolution through the semi-classical Einstein equations
\begin{equation}\label{Eq. Einstein Equations}
	R_{ab} - \frac{1}{2}R \, g_{ab} + \Lambda \, g_{ab} = M^{-2} \big\langle \hat{T}_{ab}\big\rangle 
\end{equation}
where $\Lambda$ represents the cosmological constant, ${M = ({8 \pi G})^{-1/2}}$ stands for the reduced Planck mass, and $\big\langle \hat{T}_{ab}\big\rangle$ corresponds to the expectation value of the energy-momentum tensor operator, including contributions both from the scalar field we are considering plus any other stress-energy sources. The canonical expression for $\hat{T}_{ab}$ due to the scalar field is constructed by varying the action in Eq.~(\ref{Eq. Field Action}) with respect to the metric $g_{ab}$, and subsequently substituting the field operator $\hat{\Phi}$ from Eq.~(\ref{Eq. Field}) into the resulting expression:
\begin{equation}\label{Eq. Energy-Momentum Tensor}
	\hat{T}_{ab} = \big( \nabla_{a} \hat{\Phi} \big) \big( \nabla_{b} \hat{\Phi} \big) - \frac{1}{2}g_{ab} \big( \nabla^{c} \hat{\Phi} \big) \big( \nabla_{c} \hat{\Phi} \big) + \xi \bigg[  g_{ab} \, \Box - \nabla_{a}\nabla_{b} + R_{ab} - \frac{1}{2}R \, g_{ab} - \frac{m^{2}}{2}g_{ab} \bigg] \hat{\Phi}^{2} \, .
\end{equation}
For a statistically homogeneous and isotropic field state, $\big\langle \hat{T}_{ab}\big\rangle$ is equivalent to the energy-momentum tensor of a perfect fluid for which the field energy density and pressure are given, respectively, by $\rho(t) = \big\langle \hat{T}_{00} \big\rangle$ and $P(t) = \frac{1}{3}g^{ij}\big\langle \hat{T}_{ij} \big\rangle$. As a consequence, such a state naturally sources an FLRW metric, reducing Eq.~(\ref{Eq. Einstein Equations}) to the usual Friedmann equations:
\begin{subequations}\label{Eq. Friedmann Equations}
\begin{align}
	& H^{2}(t) = \frac{1}{3} M^{-2} \rho(t) + \frac{\Lambda}{3} - \frac{K}{a^{2}(t)} \label{Eq. First Friedmann Equation} \\[8pt]
	&\dot{H}(t) + H^{2}(t) = -\frac{1}{6} M^{-2} \Big[ \rho(t) + 3P(t) \Big] + \frac{\Lambda}{3} \, .
\end{align}
\end{subequations}
Furthermore, it can be shown that $\big\langle \hat{T}_{ab}\big\rangle$ is covariantly conserved, resulting in the cosmological continuity equation
\begin{equation}\label{Eq. Cosmological Continuity}
	\dot{\rho}(t) + 3H(t)\Big[ \rho(t) + P(t) \Big] = 0 \, .
\end{equation}

However, at this stage these equations are merely formal, because both the pressure and energy density of a quantized field are in general divergent and need to be regularized. 
In a FLRW spacetime, these divergencies can be partially isolated by expressing $\rho(t)$ and $P(t)$ in terms of the adiabatic bilinears $\mathcal{N}_{k}(t)$ and $\mathcal{M}_{k}(t)$ \cite{Habib99}. Substituting Eqs.~(\ref{Eq. Field}), (\ref{Eq. Adiabatic Bogolyubov Mode}), and (\ref{Eq. Adiabatic Bogolyubov Mode Derivative}) into the expectation value of Eq.~(\ref{Eq. Energy-Momentum Tensor}) and collecting terms with the same adiabatic factor gives
\begin{subequations}\label{Eq. Energy Density and Pressure}
\begin{align}
	&\rho(t) = \big\langle \hat{T}_{00} \big\rangle = \frac{1}{4 \pi a^{3}(t)} \! \int d\mu(k) \Bigg\{  \rho^{\mathcal{N}}_{k}\!(t) \bigg[ \mathcal{N}_{k}(t) + \frac{1}{2} \bigg] + \rho^{\mathcal{R}}_{k}(t)\,\mathcal{R}_{k}(t) + \rho^{\mathcal{I}}_{k}(t) \, \mathcal{I}_{k}(t) \Bigg\} \label{Eq. Energy Density} \\[8pt] 
	& P(t) = \frac{1}{3} g^{ij} \big\langle \hat{T}_{ij} \big\rangle = \frac{1}{4 \pi a^{3}(t)} \! \int d\mu(k) \Bigg\{  P^{\mathcal{N}}_{k}\!(t) \bigg[ \mathcal{N}_{k}(t) + \frac{1}{2} \bigg] + P^{\mathcal{R}}_{k}(t)\,\mathcal{R}_{k}(t) + P^{\mathcal{I}}_{k}(t) \, \mathcal{I}_{k}(t) \Bigg\} \label{Eq. Pressure} \, .
\end{align}
\end{subequations}
The terms proportional to $\mathcal{N}_{k}(t)$ capture the contribution to the energy density and pressure due to the evolving distribution of adiabatic particles populating the field modes, while the quantum interference terms contain ${\mathcal{R}_{k}(t) = \mathrm{Re}\big[ \mathcal{M}_{k}(t) \big]}$ and ${\mathcal{I}_{k}(t) = \mathrm{Im}\big[ \mathcal{M}_{k}(t) \big]}$ \footnote{Here the definitions for $\mathcal{R}_{k}(t)$ and $\mathcal{I}_{k}(t)$ might seem to differ from those found in the literature by a phase factor, but this factor is implicit in our definitions for $\alpha_{k}(t)$ and $\beta_{k}(t)$ obtained from asymptotic analysis}. The prefactors in each of these terms are defined as
\begin{subequations}\label{Eq. Adiabatic Factors}
\begin{align}
	&\rho^{\mathcal{N}}_{k}\!(t) \equiv \frac{1}{W_{k}(t)} \Bigg\{ W^{2}_{k}(t) + \omega^{2}_{k}(t) + \frac{1}{4} \Big[ V_{k}(t) - H(t) \Big]^{\!2} + \big(6\xi - 1\big) \bigg[ H(t) V_{k}(t) - 2H^{2}(t) + \frac{K}{a^{2}(t)} \bigg] \Bigg\}\, , \\[8pt]
	&\rho^{\mathcal{R}}_{k}(t) \equiv \frac{1}{W_{k}(t)} \Bigg\{ - W^{2}_{k}(t) + \omega^{2}_{k}(t) + \frac{1}{4} \Big[ V_{k}(t) - H(t) \Big]^{\!2} + \big(6\xi - 1\big) \bigg[ H(t) V_{k}(t) - 2H^{2}(t) + \frac{K}{a^{2}(t)} \bigg] \Bigg\}\, , \\[8pt]
	&\rho^{\mathcal{I}}_{k}(t) \equiv  V_{k}(t) - H(t) +2H(t)\big(6\xi - 1\big)\, , \\[8pt]
	&P^{\mathcal{N}}_{k}\!(t) \equiv \frac{1}{3W_{k}(t)} \Bigg\{ W^{2}_{k}(t) + \omega^{2}_{k}(t) - 2m^{2} + \frac{1}{4} \Big[ V_{k}(t) - H(t) \Big]^{\!2} + \frac{1}{3}\big(6\xi - 1\big)^{\!2}R(t) + \Bigg. \, \\
	 & \quad \quad \quad \quad \quad \quad \quad \quad \Bigg. + \big(6\xi - 1\big) \bigg[ -2W^{2}_{k}(t) - \frac{1}{2}V^{2}_{k}(t) + 4H(t)V_{k}(t) + 2\omega^{2}_{k}(t) + 2\dot{H}(t) + \frac{K}{a^{2}(t)} - \frac{5}{2}H^{2}(t) \bigg] \Bigg\}\, , \nonumber \\[8pt]
	&P^{\mathcal{R}}_{k}(t) \equiv \frac{1}{3W_{k}(t)} \Bigg\{ -W^{2}_{k}(t) + \omega^{2}_{k}(t) - 2m^{2} + \frac{1}{4} \Big[ V_{k}(t) - H(t) \Big]^{\!2} + \frac{1}{3}\big(6\xi - 1\big)^{\!2}R(t) + \Bigg.\,  \\
	 & \quad \quad \quad \quad \quad \quad \quad \quad \Bigg. + \big(6\xi - 1\big) \bigg[ 2W^{2}_{k}(t) - \frac{1}{2}V^{2}_{k}(t) + 4H(t)V_{k}(t) + 2\omega^{2}_{k}(t) + 2\dot{H}(t) + \frac{K}{a^{2}(t)} - \frac{5}{2}H^{2}(t) \bigg] \Bigg\}\, , \nonumber \\[8pt]
	&P^{\mathcal{I}}_{k}(t) \equiv \frac{1}{3}\Big[ V_{k}(t) - H(t) \Big] + \frac{2}{3} \big(6\xi - 1\big) \Big[ 4H(t) - V_{k}(t) \Big] \, .
\end{align}
\end{subequations}
For adiabatic field states, the contributions to the energy density and pressure due to the real bilinears $\mathcal{N}_{k}(t)$, $\mathcal{R}_{k}(t)$, and $\mathcal{I}_{k}(t)$ are always finite. This implies that the divergencies in Eqs.~(\ref{Eq. Energy Density}) and (\ref{Eq. Pressure}) are isolated in the vacuum-like terms characterized by the $\frac{1}{2}$ factors, henceforth identified as
\begin{subequations}\label{Eq. Energy and Pressure Vacuum}
\begin{align}
	\rho_{\mathrm{vac}}(t) &\equiv \frac{1}{8 \pi a^{3}(t)} \! \int d\mu(k)\,\rho^{\mathcal{N}}_{k}\!(t) \label{Eq. Vacuum Energy} \\[8pt]
	P_{\mathrm{vac}}(t) &\equiv \frac{1}{8 \pi a^{3}(t)} \! \int d\mu(k)\,P^{\mathcal{N}}_{k}\!(t) \label{Eq. Vacuum Pressure} \, .
\end{align}
\end{subequations}

Regularization consists precisely in controlling the divergent behavior of $\rho_{\mathrm{vac}}(t)$ and $P_{\mathrm{vac}}(t)$ so as to obtain finite expressions for  $\rho(t)$ and $P(t)$ which still satisfy the cosmological continuity equation. Adiabatic regularization achieves this result by subtracting the fourth-order phase-integral expansions of $\rho^{\mathcal{N}}_{k}\!(t)$ and $P^{\mathcal{N}}_{k}\!(t)$ from the integrands of Eqs.~(\ref{Eq. Vacuum Energy}) and (\ref{Eq. Vacuum Pressure}), respectively \cite{ParkerFulling74, AndersonParker87}. From a technical perspective, however, this procedure introduces significant challenges to the numerical implementation of the semi-classical Friedmann equations. Chief among these is the appearance of higher-order time derivatives of $H(t)$ in the integrands of Eqs.~(\ref{Eq. Energy Density and Pressure}), turning the semi-classical Friedmann equations into a system of integro-differential equations which is not amenable to standard numerical treatments. We circumvent this difficulty by employing an alternative regularization scheme which, albeit cruder, yields a good approximation to $\rho(t)$ and $P(t)$ in regimes dominated by particle production.

Central to the regularization approach adopted here is the realization that $\rho_{\mathrm{vac}}(t)$ and $P_{\mathrm{vac}}(t)$ independently satisfy the cosmological continuity equation as long as the function $V_{k}(t)$ has the form established in Eq.~(\ref{Eq. Adiabatic V}) (see the Appendix). It follows that the vacuum contributions to $\rho(t)$ and $P(t)$ can be discarded in their entirety while still ensuring that Eq.~(\ref{Eq. Cosmological Continuity}) remains valid. Despite its simplicity, this procedure yields a good approximation to the field energy density and pressure provided the adiabatically regularized integrands of Eqs.~(\ref{Eq. Energy Density and Pressure}) are dominated by the real adiabatic bilinears $\mathcal{N}_{k}(t)$, $\mathcal{R}_{k}(t)$, and $\mathcal{I}_{k}(t)$. Therefore, in what follows we take the regularized expressions for the energy density and pressure to be
\begin{subequations}\label{Eq. Reg. Energy Density and Pressure}
\begin{align}
	\rho(t) &\approx \frac{1}{4 \pi a^{3}(t)} \! \int d\mu(k) \Bigg\{  \rho^{\mathcal{N}}_{k}\!(t) \, \mathcal{N}_{k}(t) + \rho^{\mathcal{R}}_{k}(t)\,\mathcal{R}_{k}(t) + \rho^{\mathcal{I}}_{k}(t) \, \mathcal{I}_{k}(t) \Bigg\} \label{Eq. Reg. Energy Density} \\[8pt] 
	P(t) &\approx \frac{1}{4 \pi a^{3}(t)} \! \int d\mu(k) \Bigg\{  P^{\mathcal{N}}_{k}\!(t) \, \mathcal{N}_{k}(t) + P^{\mathcal{R}}_{k}(t)\,\mathcal{R}_{k}(t) + P^{\mathcal{I}}_{k}(t) \, \mathcal{I}_{k}(t) \Bigg\} \label{Eq. Reg. Pressure} \, ,
\end{align}
\end{subequations}
where the factors $\rho^{\mathcal{N}}_{k}\!(t)$, $\rho^{\mathcal{R}}_{k}(t)$, and $\rho^{\mathcal{I}}_{k}(t)$, $P^{\mathcal{N}}_{k}\!(t)$, $P^{\mathcal{R}}_{k}(t)$, and $P^{\mathcal{I}}_{k}(t)$ are computed by truncating the asymptotic series Eq.~(\ref{Eq. Adiabatic W}) for $W_{k}(t)$ at its optimal order.

Finally, the regularization of $\big\langle \hat{T}_{ab}\big\rangle$ also induces the renormalization of the gravitational coupling constants $G$ and $\Lambda$. Moreover, self-consistency demands the  introduction of a covariantly conserved tensor composed of fourth-order derivatives of the metric into the semi-classical Einstein equations \cite{FullingParker74, Bunch80}. This tensor is accompanied by a new unknown coupling constant whose renormalization assimilates the ultra-violet divergence in the field energy-momentum tensor. For simplicity, in this work we assume this new coupling constant to be renormalized to zero, thus preserving the form of Eq.~(\ref{Eq. Einstein Equations}). Non-zero values for this coupling constant will be considered elsewhere.

Taken together, Eqs.~(\ref{Eq. Bilinears Multiple Crossings}), (\ref{Eq. Friedmann Equations}), and (\ref{Eq. Reg. Energy Density and Pressure}) describe the coupled field evolution and cosmic evolution in regimes dominated by particle production. In the next section we present an algorithm which numerically solves this system of equations.\\
\twocolumngrid

\section{Numerical Implementation}\label{Sec. Numerical Implementation}

\begin{figure}[t!]
    \includegraphics[width=0.49\textwidth]{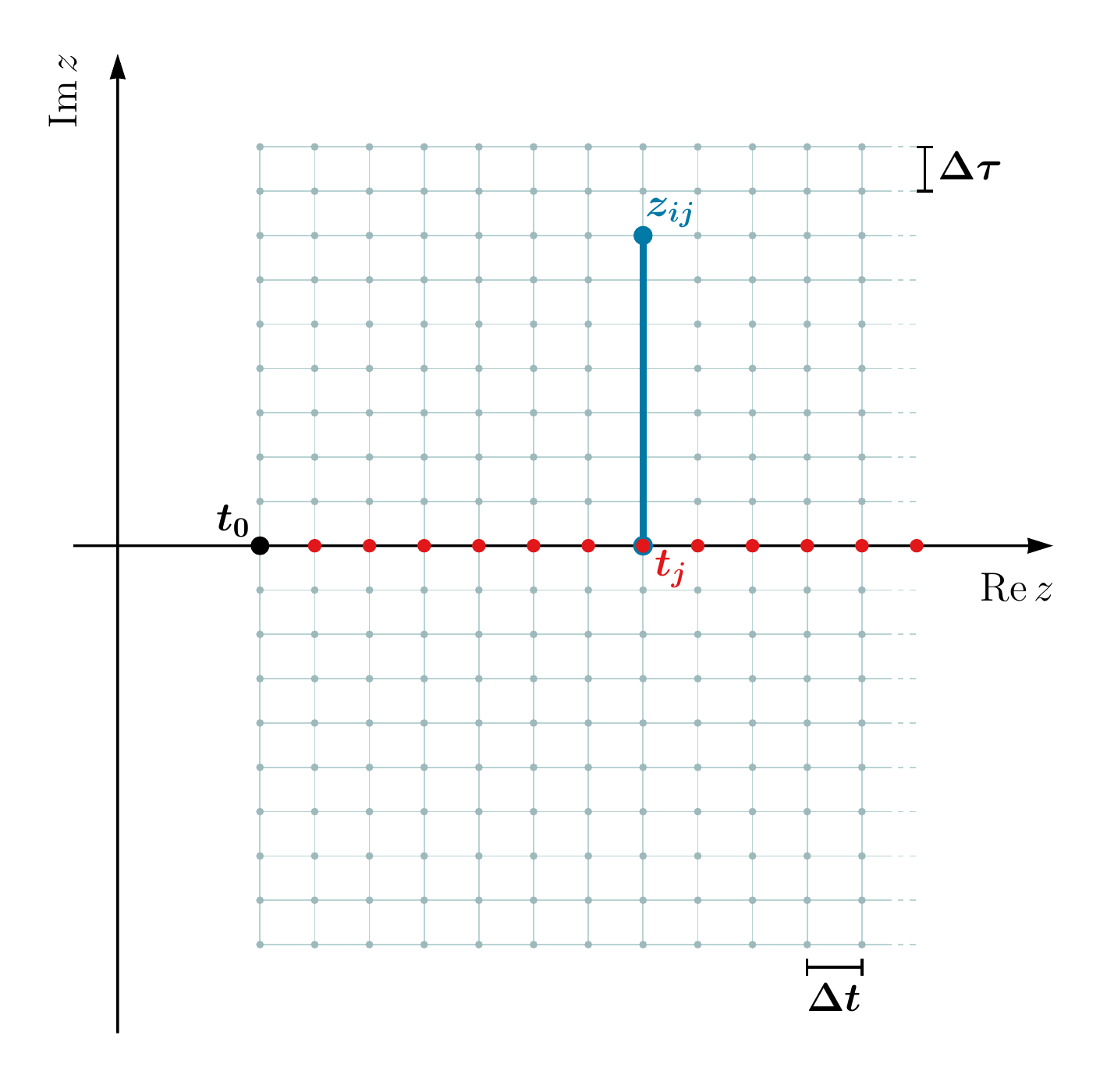}
    \caption{A grid of uniformly spaced points covering a band of the complex plane. The grid points lying over the real axis mark the discretization of physical time. Numerically constructing the Taylor polynomial associated with the frequency function around the point $t_{j}$ allows for the optimal evaluation of $\Omega_{k}(z_{ij})$ at grid points $z_{ij}$ for which $\mathrm{Re}\,z_{ij} = t_{j}$.}
    \label{Fig. Numerical Analytic Continuation}
\end{figure}

The semi-classical Friedmann equations can be formulated as a discretized initial value problem. We take the domain of numerical integration to be a band of the complex plane which is bisected by the real $t$ axis. As illustrated in Figure \ref{Fig. Numerical Analytic Continuation}, this band is discretized by a uniformly spaced grid where the real-valued entries $t_{j}$ label the physical time. Initial conditions are set by an appropriately chosen functional form for the scale factor $a(t)$ which not only admits an adiabatic field state at the initial time $t_{0}$, but which is also consistent with our choice for the initial distribution of adiabatic particles $\mathcal{N}_{k}(t_{0}) = N_{k}$ populating the field modes. In addition, we require that
\begin{equation}\label{Eq. Particle Distribution Constraint}
	N_{k} < \mathcal{O}(k^{-3}) \quad \text{as} \quad k \rightarrow \infty
\end{equation}
in order to ensure that both the energy density and pressure associated with the initial particle distribution are finite. 

We use a standard finite-difference scheme to step $a(t)$, $H(t)$, and $\dot{H}(t)$ along the real axis, and employ B-splines to scan the Stokes geometry on the complex plane. The latter is accomplished by generating a numerical sample of $\Omega_{k}(t)$ through Eq.~(\ref{Eq. Mode Frequency}), and subsequently performing high-order B-spline interpolations to construct a truncated Taylor polynomial for this function over the real line up to the value of $t$ in the current time step. Due to the analyticity of $\Omega_{k}(t)$, this series representation is also valid on the complex plane, and thus encodes the analytical continuation of the frequency function. Explicitly, given a grid point $z_{ij}$ on the discretized plane, we compute $\Omega_{k}(z_{ij})$ through the expression
\begin{equation}\label{Eq. Taylor}
\Omega_{k}(z_{ij}) \approx \sum_{n = 0}^{T} \frac{1}{{n!}} \big(z_{ij} - t_{j} \big)^{n} \, \Omega_{k}^{(n)}(t_{j}) \, ,
\end{equation}
where $t_{j} = \mathrm{Re}\,z_{ij}$, as depicted in Figure \ref{Fig. Numerical Analytic Continuation}. The numerical derivatives $\Omega_{k}^{(n)}$ are extracted from B-spline interpolations over the real axis, and $T$ corresponds to a truncation order which depends on the density of grid points lying over the real axis. In addition, we feed Eq.~(\ref{Eq. Taylor}) to a Pad\'{e} approximant \cite{Bender13} routine to accelerate its convergence and improve its accuracy. Once this approximate representation of the frequency function has been computed over the discretized plane, it can be interpolated and used in the monitoring of turning points and Stokes lines.

While the turning points of $\Omega_{k}(z)$ can be located with the aid of root-finding algorithms designed for multi-valued functions, the problem of determining the Stokes lines sourced by these points requires the numerical integration of an ordinary differential equation. This is evident from the Stokes lines definition ${\mathrm{Re}\big[\Omega_{k}\mathrm{d}z\big]=0}$, which implies that, locally, its line element must satisfy ${\mathrm{d}z \propto i/\Omega_{k}(z)}$. Defining ${t=\mathrm{Re} \, z}$ and ${\tau=\mathrm{Im} \, z}$, this condition can be rewritten as
\begin{equation}
	\mathrm{d}z = \mathrm{d}t + i\,\mathrm{d}\tau \propto \frac{i}{\Omega_{k}(z)} \, .
\end{equation}
Taking the ratio between the matched real and imaginary parts of this proportionality relation leads to the differential equation 
\begin{equation}\label{Eq. Stokes ODE}
	\frac{dt}{d\tau} = \frac{ \mathrm{Im} \, \Omega_{k}(z) }{ \mathrm{Re} \, \Omega_{k}(z) } \, 
\end{equation}
for the Stokes line,
which can be numerically integrated from the turning point of interest to yield $t(\tau)$.

Here is a summary of the minimal set of tasks performed by our algorithm while evolving the physical quantities of interest by one time step:
\begin{itemize}
	\item[1.] Take samples of $a(t)$, $H(t)$, and $\dot{H}(t)$ describing the metric along an interval of the real axis. Over this same interval, sample and interpolate the field energy density $\rho(t)$ and pressure $P(t)$.
	
	\item[2.] Numerically integrate the semi-classical Friedmann equations so as to enlarge the  input metric samples $a(t)$, $H(t)$, and $\dot{H}(t)$ by a time step $\Delta t$.

	\item[3.] For each field mode, generate a sample of the frequency function $\Omega_{k}(t)$ over the real axis, and numerically extend this function onto the complex plane to obtain $\Omega_{k}(z)$.
	
	\item[4.] Search for complex turning points of each frequency function $\Omega_{k}(z)$, and numerically trace their corresponding Stokes lines.

	\item[5.] If a Stokes line associated with a mode of wavenumber $k$ is found to intersect the real axis, update the real bilinears $\mathcal{N}_{k}(t)$, $\mathcal{R}_{k}(t)$ and $\mathcal{I}_{k}(t)$ accordingly.
	
	\item[6.] For each field mode, compute $W_{k}(t)$ and $V_{k}(t)$ up to the optimal truncation order set by the last Stokes line crossing.
	
	\item[7.] Gather the results from all previous steps to evolve the input samples for the field energy density $\rho(t)$ and pressure $P(t)$ by a time step $\Delta t$.
\end{itemize}

In general, the Stokes lines associated with field modes of comparable wavenumber will cross the real axis within close proximity of one another, giving rise to overlapping particle production events. In order to correctly capture the influence that such events might have on each other, we apply the stepping algorithm outlined above in an iterative fashion. In other words, once the quantities of interest have been forward-stepped up to $t_{j}$, the following iteration backtracks to $t_{0}$ and then proceeds to step the problem up to $t_{j+1} = t_{j} + \Delta t$ using as sources for the semi-classical Friedmann equations the field energy density and pressure obtained in the previous iteration.

In summary, our numerical implementation allows for the scale factor and the Stokes geometry to reconfigure themselves with each iteration and thereby construct a self-consistent solution to the back-reaction problem. 

\section{Numerical Results}\label{Sec. Results}

To assess the accuracy of our numerical approach, we first neglect back-reaction effects and compare numerical results to known analytic solutions for a quantized scalar field evolving in a closed de Sitter spacetime \cite{Mottola85}. This case is characterized by a positive cosmological constant $\Lambda$ and a curvature parameter of $K = 1$, which together lead to a bouncing scale factor evolution
\begin{equation}\label{Eq. de Sitter}
	a(t) = \bar{H}^{-1} \cosh{\left(\bar{H} t\right)} \quad \text{with} \quad \bar{H} = \sqrt{\Lambda/3} \, .
\end{equation}
Here $\bar{H}$ is the asymptotic value of the Hubble parameter in the infinite future,
\begin{equation}\label{Eq. Hubble de Sitter}
	\lim_{t \to \pm \infty}{H(t)} = \pm \bar{H} \, .
\end{equation}
This model Universe contracts for $t < 0$, reaches its minimum size at $t = 0$, and subsequently expands for the $t > 0$.

Substituting Eq.~(\ref{Eq. de Sitter}) into Eq.~(\ref{Eq. Mode Frequency}) yields 
\begin{equation}\label{Eq. de Sitter Mode Frequency}
\Omega_{k}^{2}(t) = \bar{H}^{2} \Bigg[ \bigg(k^{2} - \frac{1}{4} \bigg) \, \text{sech}^{2} \! \left(\bar{H} t\right) + \frac{m^{2}}{\bar{H}^{2}} + 12\,\xi - \frac{9}{4} \Bigg] \, 
\end{equation}
for the mode frequency function.
Analytically extending this function to the complex plane, locating its turning points, and tracing its Stokes lines are straightforward. We verify our numerical calculations against these analytic results. For definiteness, we choose a scalar field of mass $m = 0.1\,M$ which is conformally coupled to the scalar curvature, $\xi = \frac{1}{6}$. We set the cosmological constant to $\Lambda = 3 \, m^{2}$, so that $\bar{H} = 1 \, m$. All dimensional quantities are thus expressed in terms of the field mass.

\begin{figure}[t!]
    \includegraphics[width=0.49\textwidth]{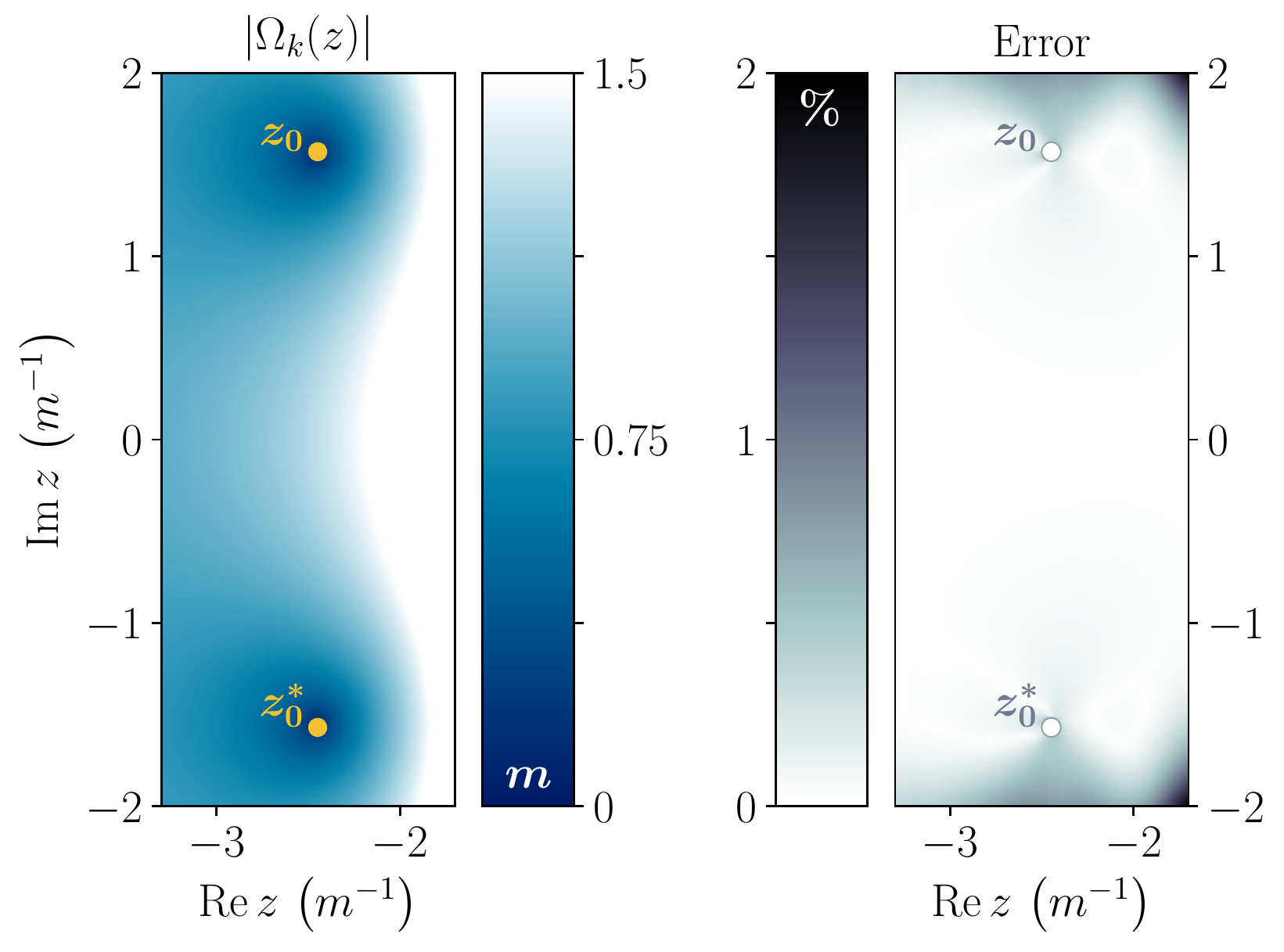}
    \caption{A comparison between the numerical analytic continuation of $\Omega_{k}(z)$ produced by our algorithm and the expected analytic expression for this function in a closed de Sitter spacetime. The field parameters are $m = 0.1 \, M$, $\xi = \frac{1}{6}$, and $k = 5 \, m$, while the spacetime is characterized by $\Lambda = 3 \, m^{2}$ and $K = 1$. The left panel shows the absolute value of the numerically produced frequency function in the vicinity of the pair of conjugate turning points $\big( z_{0},\, z_{0}^{\ast}\big)$, while the right panel exhibits the relative difference between the analytic and numerical results.}
    \label{Fig. Continuation Error}
\end{figure}

\begin{figure*}[hbtp]
    \includegraphics[width=1.0\textwidth]{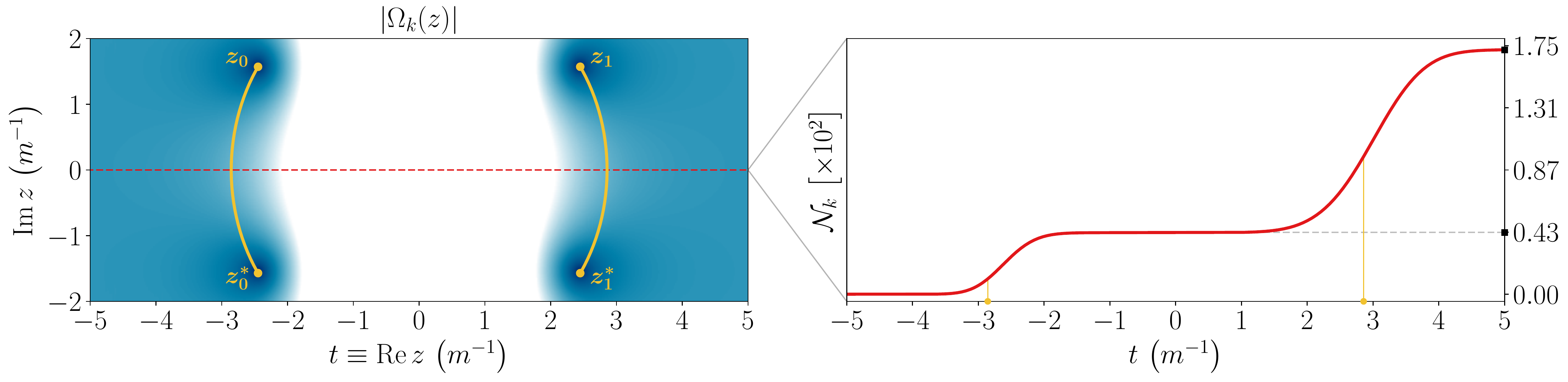}
    \caption{The numerically traced Stokes geometry associated with the frequency function $\Omega_{k}(z)$, and the adiabatic particle number evolution $\mathcal{N}_{k}(t)$ extracted from it. The field parameters are set to $m = 0.1 \, M$, $\xi = \frac{1}{6}$, $k = 5 \, m$, and $N_{k} = 0$, while the spacetime is characterized by $\Lambda = 3 \, m^{2}$ and $K = 1$. The left panel shows the Stokes lines sourced by the pairs of turning points $\big( z_{0},\, z_{0}^{\ast}\big)$ and $\big( z_{1},\, z_{1}^{\ast}\big)$ superimposed over the absolute value of the numerically obtained frequency function. The real axis corresponds to the central dashed line. The effects of each Stokes line on the adiabatic particle number $\mathcal{N}_{k}(t)$ are illustrated on the right panel, wherein this quantity is tracked as a function of time. Each burst of particle production is prompted by a Stokes line crossing, indicated here by the circular markers on the horizontal axis. The expected values for the particle number plateaus featuring in this image are indicated by the square markers on the vertical axis, both of which show very good agreement with the numerically produced curve for $\mathcal{N}_{k}(t)$. Constructive interference causes more particles to be produced in the second burst.}
    \label{Fig. Frequency Particle}
\end{figure*}

A comparison between the analytic extension of Eq.~(\ref{Eq. de Sitter Mode Frequency}) and the  numerical analytic continuation produced by our algorithm is displayed in Figure \ref{Fig. Continuation Error} for the field mode of wavenumber $k = 5 \, m$. The left panel shows the absolute value of the numerically obtained frequency function, while the right panel exhibits how this result deviates from the analytic expression for $\Omega_{k}(z)$. In addition to correctly reproducing the function's conjugate pair of zeroes $\big( z_{0},\, z_{0}^{\ast}\big)$ located in this region, the numerical analytic continuation differs from the analytic value by at most 2\% in the vicinity of these points. As a result, the Stokes lines which occupy this area of the complex plane can be traced with high fidelity. This is demonstrated in the left panel of Figure \ref{Fig. Frequency Particle}, where the Stokes lines sourced by the pairs of turning points $\big( z_{0},\, z_{0}^{\ast}\big)$ and $\big( z_{1},\, z_{1}^{\ast}\big)$ are superimposed over the numerically obtained frequency function. The effects of each Stokes line on the adiabatic bilinear $\mathcal{N}_{k}(t)$ are displayed in the right panel of Figure \ref{Fig. Frequency Particle}, wherein this quantity is tracked as a function of time. Each burst of particle production is prompted by a Stokes line crossing, the first of which occurs as the Universe contracts and the field mode under consideration becomes sub-horizon; while the second burst happens after the bounce, when the mode reverts back to being super-horizon due to the Universe's expansion \cite{Habib99}. Despite the symmetry between these events, constructive interference expressed by Eq.~(\ref{Eq. Bilinears Multiple Crossings}) causes more particles to be produced in the second burst. The expected values for the particle number plateaus are indicated by the square markers on the vertical axis, both of which agree well with the numerical curve.

\begin{figure*}[hbtp]
    \includegraphics[width=1.0\textwidth]{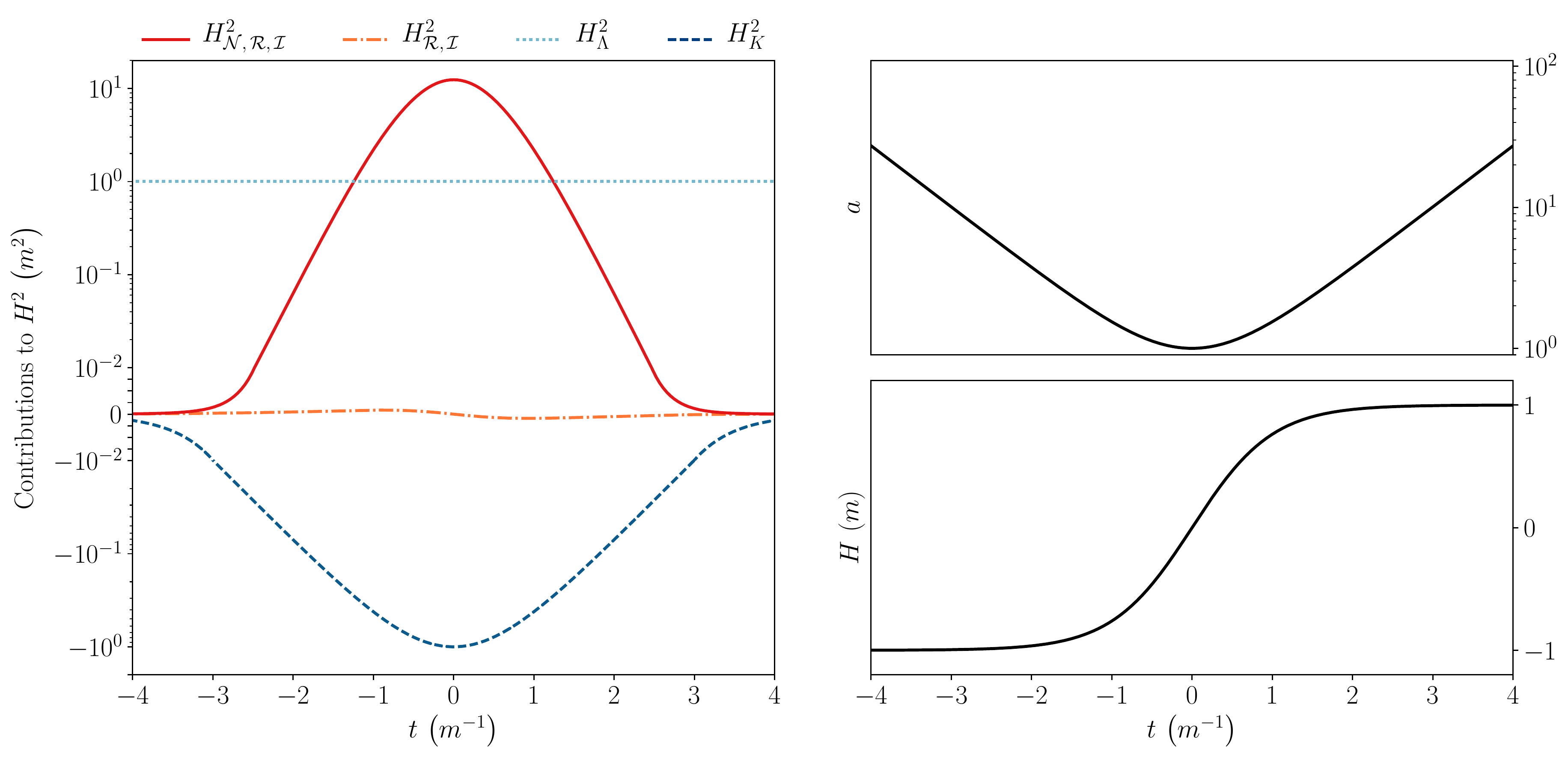}
    \caption{The evolution of every term appearing on the right-hand side of the semi-classical Friedmann Eq.~(\ref{Eq. First Friedmann Equation}) in a closed de Sitter spacetime evolution, as well as the quantities describing the metric for this spacetime in the absence of back reaction. The field parameters are set to $m = 0.1 \, M$ and $\xi = \frac{1}{6}$, while the spacetime is characterized by $\Lambda = 3 \, m^{2}$ and $K = 1$. The bounce starts at $t_{0} = -5 \, m^{-1}$ with an initial particle distribution given by $\mathcal{N}_{k}(t_{0}) = 0$. The left panel follows the evolution of $H^{2}_{\mathcal{N},\,\mathcal{R},\,\mathcal{I}}$ (solid line), $H^{2}_{\mathcal{R},\,\mathcal{I}}$ (dot-dashed line), $H^{2}_{\Lambda}$ (dotted line), and $H^{2}_{K}$ (dashed line). While $H^{2}_{\mathcal{R},\,\mathcal{I}}$ remains negligible throughout, $H^{2}_{\mathcal{N},\,\mathcal{R},\,\mathcal{I}}$ grows exponentially and eventually comes to dominate over all other contributions. The right panels illustrate the scale factor $a(t)$ and Hubble parameter $H(t)$ which describe the de Sitter bounce. Because back-reaction effects are being neglected, the Hubble parameter is just $H^{2} = H^{2}_{\Lambda} + H^{2}_{K}$.} 
    \label{Fig. Contributions to H2 - No back-reaction}
\end{figure*}

By tracing the Stokes geometry of every field mode, we can also track the evolution of the field energy density as the spacetime evolves. Even though back-reaction effects are being neglected, this quantity shows whether the effects of particle production will eventually become comparable to the contributions from $\Lambda$ and $K$ which source the background de Sitter spacetime. To that end, we track every term appearing on the right-hand side of  the semi-classical Friedmann Eq.~(\ref{Eq. First Friedmann Equation}), identifying each contribution according to the notation
\begin{subequations}\label{Eq. Contributions to H2}
\begin{align*}
	H^{2}_{\mathcal{N},\,\mathcal{R},\,\mathcal{I}} \equiv \frac{\rho}{3M^2}\,\,, \quad H^{2}_{\Lambda} \equiv \frac{\Lambda}{3} \,\,, \quad \text{and} \quad H^{2}_{K} \equiv - \frac{K}{a^{2}} \, .
\end{align*}
\end{subequations}
Additionally, we define $H^{2}_{\mathcal{R},\,\mathcal{I}}$ as the contribution to the right-hand side of Eq.~(\ref{Eq. First Friedmann Equation}) which stems solely from terms proportional to the real bilinears $\mathcal{R}_{k}$ and $\mathcal{I}_{k}$. The left panel of Figure~\ref{Fig. Contributions to H2 - No back-reaction} displays the evolution of the above-defined quantities for a bounce that starts at $t_{0} = -5 \, m^{-1}$ with an initial particle distribution given by $\mathcal{N}_{k}(t_{0}) = 0$. Being the only true sources in this case, $H^{2}_{\Lambda}$ and $H^{2}_{K}$ behave in the standard way, acting in concert to produce the de Sitter bounce. Because back-reaction effects are neglected, the Hubble parameter $H$ shown on the right panel of Figure~\ref{Fig. Contributions to H2 - No back-reaction} is entirely characterized by these two quantities, i.e., $H^{2} = H^{2}_{\Lambda} + H^{2}_{K}$. On the other hand, the field-related quantities $H^{2}_{\mathcal{N},\,\mathcal{R},\,\mathcal{I}}$ and $H^{2}_{\mathcal{R},\,\mathcal{I}}$ display an interesting behavior which mirrors the result found in Ref.~\cite{AndersonMottola14}. While $H^{2}_{\mathcal{R},\,\mathcal{I}}$ remains negligible throughout, $H^{2}_{\mathcal{N},\,\mathcal{R},\,\mathcal{I}}$ grows exponentially as the Universe progresses toward the bounce. In other words, the field energy density eventually becomes dominated by $\mathcal{N}_{k}$ -- the field particle content. Physically, the soaring field energy density is due to the blueshift experienced by particles produced in the contracting phase. As a result, the Universe is filled with relativistic particles which effectively behave as radiation, making the field energy density grow as $\rho \propto a^{-4}$. This trend is then reversed in the ensuing expanding phase, during which the field energy density drops rapidly as particles are continuously redshifted.
\begin{figure*}[hbtp]
    \includegraphics[width=1.0\textwidth]{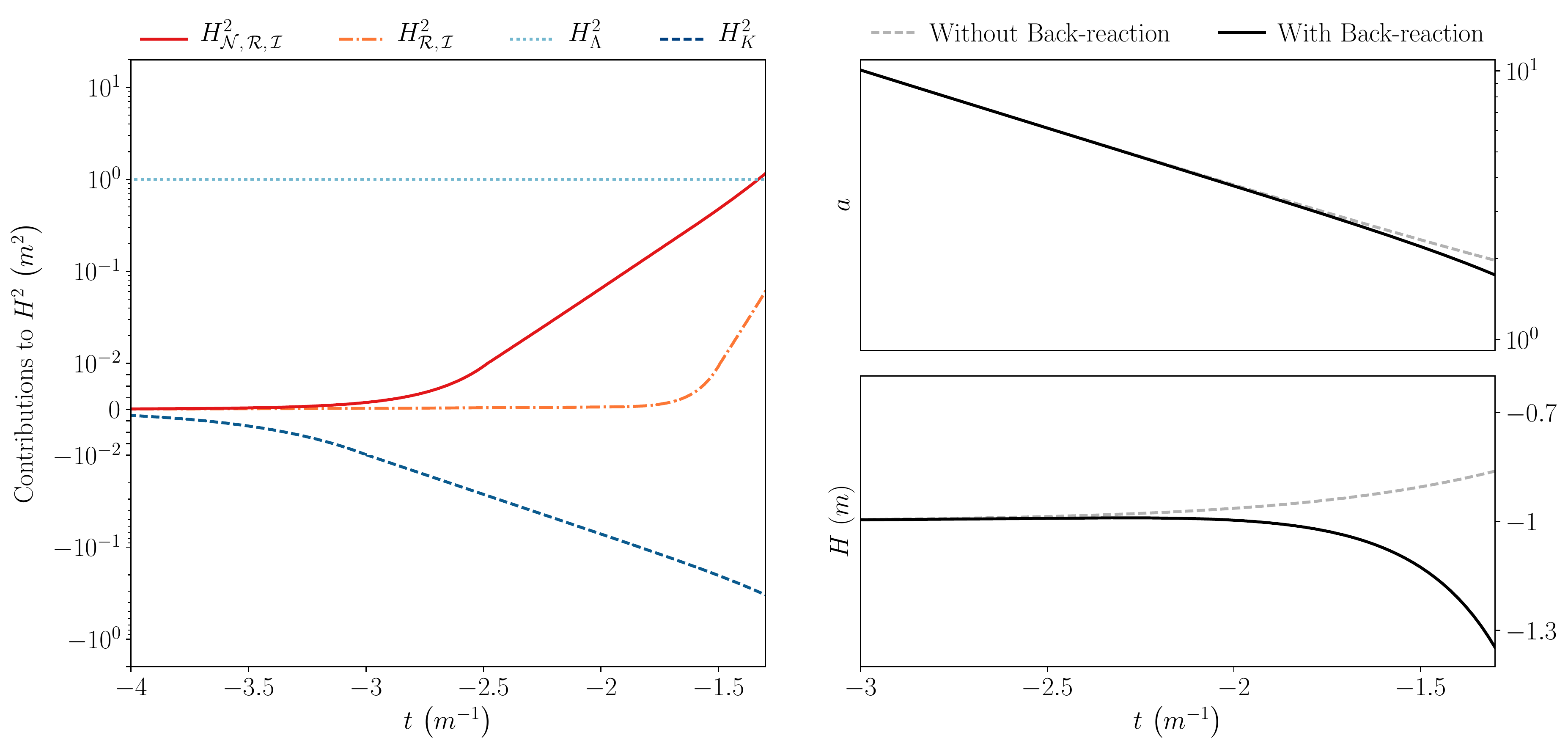}
    \caption{The evolution of every term appearing on the right-hand side of the semi-classical Friedmann Eq.~(\ref{Eq. First Friedmann Equation}) and the quantities describing the metric evolution in a full back-reacting calculation. The field parameters are $m = 0.1 \, M$ and $\xi = \frac{1}{6}$, while the cosmological constant and curvature parameter are $\Lambda = 3 \, m^{2}$ and $K = 1$. The closed de Sitter initial conditions are set at $t_{0} = -5 \, m^{-1}$, along with an initial particle distribution given by $\mathcal{N}_{k}(t_{0}) = 0$. The left panel follows the evolution of $H^{2}_{\mathcal{N},\,\mathcal{R},\,\mathcal{I}}$ (solid line), $H^{2}_{\mathcal{R},\,\mathcal{I}}$ (dot-dashed line), $H^{2}_{\Lambda}$ (dotted line), and $H^{2}_{K}$ (dashed line). The exponential growth of $H^{2}_{\mathcal{N},\,\mathcal{R},\,\mathcal{I}}$ effectively fills the Universe with relativistic particles, introducing an instability to the initial de Sitter phase. The right panels illustrate the scale factor $a(t)$ and Hubble parameter $H(t)$ transitioning from a Sitter bounce to a radiation dominated phase. Here the solid lines represent the solutions to the back-reaction problem, while the dashed lines trace the pure de Sitter bounce.} 
    \label{Fig. Contributions to H2 - Back-reaction}
\end{figure*}

The preceding calculations demonstrate that back-reaction effects due to particle production can become dynamically significant in an initially closed de Sitter spacetime. A full account of these effects is shown in Figure~\ref{Fig. Contributions to H2 - Back-reaction}, using the algorithm for computing back-reaction effects described in the previous Section. In this case, the metric evolution initially matches that of a closed de Sitter spacetime at $t_{0} = -5 \, m^{-1}$, while the initial particle distribution is given by $\mathcal{N}_{k}(t_{0}) = 0$. These initial conditions self-consistently satisfy the semi-classical Friedmann equations at the initial time $t_{0}$ within our approximations. As in the case without back reaction, the quantity $H^{2}_{\mathcal{R},\,\mathcal{I}}$ remains sub-dominant throughout the evolution, while $H^{2}_{\mathcal{N},\,\mathcal{R},\,\mathcal{I}}$ grows exponentially as newly-created particles are continuously blueshifted. Since they quickly become relativistic, these particles behave as an additional radiation-like component, destabilizing the initial de Sitter phase. This is illustrated in the right panels of Figure~\ref{Fig. Contributions to H2 - Back-reaction}, where the scale factor and Hubble parameter can be seen transitioning from a de Sitter bounce to a radiation-dominated behavior. The contributions from the regularized vacuum terms discarded in our approximations remain negligible at all times. We stop the numerical integration at $t = -1.3 \, m^{-1}$, since beyond this time the Hubble parameter becomes of order $H \simeq M^{-1}$, invalidating the semi-classical picture of gravity on which our calculations rely.

\begin{figure*}[hbtp]
    \includegraphics[width=1.0\textwidth]{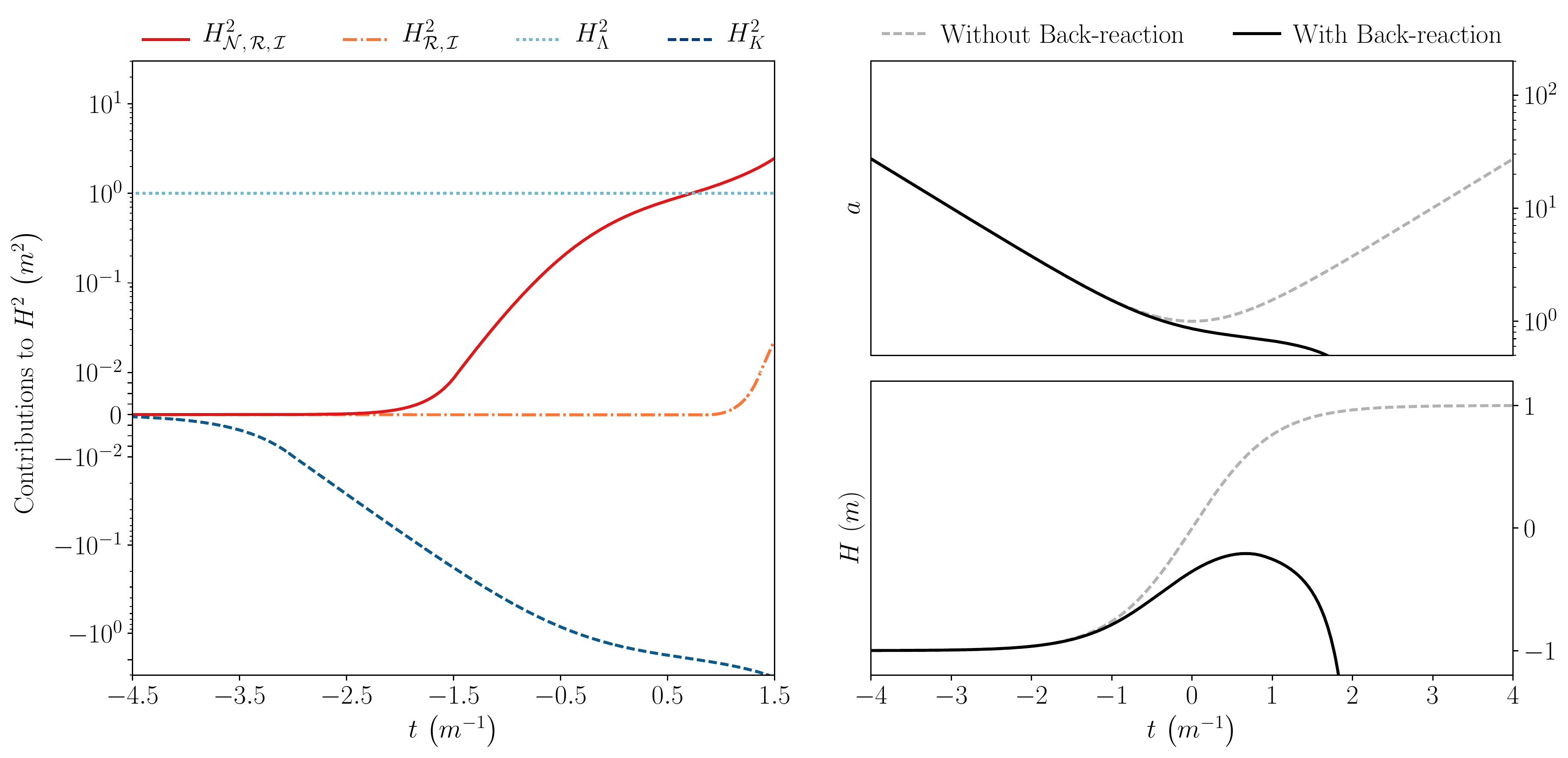}
    \caption{The evolution of every source term featuring on the right-hand side of the semi-classical Friedmann Eq.~(\ref{Eq. First Friedmann Equation}) and the quantities describing the metric evolution in a full back-reacting calculation. The field parameters are set to $m = 0.0145 \, M$ and $\xi = \frac{1}{6}$, while the cosmological constant and curvature parameter are characterized by $\Lambda = 3 \, m^{2}$ and $K = 1$. The closed de Sitter initial conditions are set at $t_{0} = -5 \, m^{-1}$, along with an initial particle distribution given by $\mathcal{N}_{k}(t_{0}) = 0$. The left panel follows the evolution of $H^{2}_{\mathcal{N},\,\mathcal{R},\,\mathcal{I}}$ (solid line), $H^{2}_{\mathcal{R},\,\mathcal{I}}$ (dot-dashed line), $H^{2}_{\Lambda}$ (dotted line), and $H^{2}_{K}$ (dashed line). The growth of $H^{2}_{\mathcal{N},\,\mathcal{R},\,\mathcal{I}}$ fills the Universe with just enough relativistic particles to destabilize the initial de Sitter phase. The right panels illustrate the scale factor $a(t)$ and Hubble parameter $H(t)$ transitioning from a de Sitter bounce to a radiation dominated phase. Here the solid lines represent the solutions to the back-reaction problem, while the dashed lines trace the pure de Sitter bounce. Had the field mass been set to a value $m \lesssim 0.0142 \, M$, a bounce would still take place, albeit at a slightly later time and for a smaller value of the scale factor.}
    \label{Fig. Contributions to H2 - Back-reaction 2}
\end{figure*}

The de Sitter bounce is not always disrupted by particle production. For sufficiently low values of the field mass, the bounce is merely delayed. Figure~\ref{Fig. Contributions to H2 - Back-reaction 2} illustrates a near-limiting case with ${m = 0.0145 \, M}$ for which an initial de Sitter evolution is still driven toward a radiation dominated phase. The contributions due to particle production $H^{2}_{\mathcal{N},\,\mathcal{R},\,\mathcal{I}}$ only come to dominate over the combined $H^{2}_{\Lambda}$ and $H^{2}_{K}$ near the bounce at $t = 0 \, m^{-1}$. For field masses $m \lesssim 0.0142 \, M$, the negative curvature contributions $H^{2}_{K}$ neutralize the growth of $H^{2}_{\mathcal{N},\,\mathcal{R},\,\mathcal{I}}$ for long enough to preserve the bounce. The resulting bounce is pushed to a slightly later time and occurs at a smaller value of the scale factor.

\section{Discussion}\label{Sec. Discussion}

The back-reaction problem addressed in this work imposes several technical hurdles which have resisted a satisfactory solution for decades. These difficulties stem primarily from the necessity to control the divergent nature of the vacuum energy. Adiabatic regularization accomplishes this at the cost of increasing the problem's complexity. As a result, ambiguities arise in the specification of initial conditions and in the value of physical quantities when particle production is rapid, and computationally the problem becomes susceptible to potential numerical instabilities. In this work we have shown that these issues can be circumvented in scenarios dominated by particle production. Our approach relies on a particular choice of adiabatic mode functions which isolate the vacuum contributions into a separate covariantly conserved component of the total stress-energy. In regimes dominated by particle production, this vacuum component is sub-dominant and can be discarded in its entirety. By definition, the remaining covariantly conserved portion of the stress-energy dominates, as it encapsulates the effects of particle production. This component can be expressed in terms of the particle number density as described by Berry's universal form, resolving the ambiguity in physical quantities, and computed from the analytic continuation of each mode's frequency function onto the complex plane (Figures \ref{Fig. Frequency Particle} and \ref{Fig. Contributions to H2 - No back-reaction}). The resulting stress-energy is a calculable source term for the semi-classical Friedmann equations, and can be used to obtain a numerical solution to the back-reaction problem. We have performed this calculation for an initially closed de Sitter spacetime, demonstrating that the effects of particle production in this scenario can become strong enough to drive the cosmic evolution into a radiation-dominated phase (Figures \ref{Fig. Contributions to H2 - Back-reaction} and \ref{Fig. Contributions to H2 - Back-reaction 2}). Our results illustrate the reliability of our numerical implementation, and open the possibility of the systematic investigation of cosmological scenarios dominated by quantum particle production.

On a technical level, our method relies on some previous knowledge of the Stokes geometry associated with the spacetime evolution. For the case studied in this work, all Stokes lines are sufficiently separated from each other so that Berry's universal form for particle production applies without corrections. In general, however, the spacetime evolution might result in near-lying Stokes lines for which higher-order Stokes corrections are required for an accurate description of particle production. Although not included in this work, such corrections are well-documented in the literature \cite{BerryHowls94, HowlsLongmanDaalhuis04, HowlsDaalhuis12} and could in principle be added to our numerical implementation. More fundamentally, our method is based on a well-defined semi-classical notion of particle. Mathematically, this notion is tied to the existence of a phase-integral expansion for the field mode functions. Such a representation can always be constructed as long as $\left| \varepsilon_{k,\,0} \right| \ll 1$. Physically, this requirement typically translates to an approximate bound on the Hubble rate $H \lesssim m$ set by the mass of the field under consideration. Nonetheless, some scenarios exist for which $\left| \varepsilon_{k,\,0} \right| \ll 1$ is satisfied even when $H > m$.

Quantum backreaction  is potentially important in models of the very early Universe. Quantum particle production is actually quite familiar in the context of inflation, as it provides the standard mechanism for the generation of perturbations in an inflating spacetime \cite{Starobinsky79, Allen88, Sahni90, Mukhanov92, Souradeep92, Glenz09, Agullo11}.  Interestingly, it has been suggested that these same ideas could be applied to the problem of driving inflation itself \cite{Prigogine89, Calvao92, Lima96, Abramo96, Gunzig98, Lima14, HaroPan16}. Indeed, a phase of accelerated expansion can result if particles are produced at a high enough rate. A time derivative of the usual Friedmann Equation $H^{2}(t) = \frac{1}{3}M^{-2} \rho(t)$ shows that an accelerating expansion ${\ddot a} > 0$ occurs when
\begin{equation}\label{Eq. Accelerated Condition}
	\dot{\rho}(t) > -\frac{2}{\sqrt{3}} \, M^{-1} \rho^{3/2}(t) \, .
\end{equation}
Such a scenario has the potential to sidestep some of the conceptual problems of the standard inflationary paradigm. For instance, it has been argued that standard inflation cannot generically start in a patch which is smaller than the cosmological horizon without violating either causality or the weak energy condition \cite{VachaspatiTrodden99, BereraGordon01}. However, if inflation is initially driven by an increasing energy density due to particle production, the weak energy condition {\it is} effectively violated. Therefore, inflation driven by such a mechanism could generically start in small patches contained within the cosmological horizon without violating causality. Inflation driven by particle production would also clarify the meaning of the inflaton effective potential by making manifest the high mass-scale physics it represents.

The same conditions which lead to quantum particle production can also result in particle annihilation. If sufficiently pronounced, this effect can drive a contracting spacetime toward a bounce phase. Indeed, it follows from the cosmological continuity equation that $\rho(t) + P(t) < 0$ provided the particle annihilation rates are high enough to cause the field energy density to decrease as the Universe contracts. In other words, the null energy condition is effectively violated, making $\dot{H}(t) > 0$ according to the Friedmann equations \cite{IjjasSteinhardt18, Ijjas16}. Thus, a classical bounce can emerge provided enough energy density is sequestered by quantum particle annihilation during a phase of cosmological contraction. If realized, such a mechanism could provide a natural description for cosmological bounce scenarios which does not require new physics. Also, successful bounces require constraints on high mass-scale quantum fields, so that quantum back-reaction does not push the contracting phase into a radiation crunch, as with the example solved in this paper.

Another possibly interesting effect is the production of a relativistic condensate in the early universe. Under certain circumstances, quantum particle production can lead to large occupation numbers for some scalar field modes, representing condensate formation. This phenomenon could lead to additional interesting phenomenology \cite{AragaoRosa80, ParkerZhang91, ParkerZhang93}. 

A number of technical questions remain to be answered. Fermion fields require more complex calculations than scalar fields, and may present some different physics \cite{Landete13, Landete14}. How to handle interacting fields remains an open question, and multiple fields offer additional possibilities \cite{Ringwald87, CooperMottola87, PazMazzitelli88, Habib96, Cooper97, MolinaParis00}. We have a long road to travel before the range of interesting early-Universe dynamical scenarios driven by quantum particle production has been fully explored.

\begin{acknowledgments}
F.Z. thanks D. Boyanovsky, S. Habib, P. Anderson, and E. Mottola for helpful discussions. F.Z. acknowledges support from the Andrew Mellon Predoctoral Fellowship and the A\&S PITT PACC Fellowship. The authors have been partly supported by the National Science Foundation under the grant AST-1312380. This work made use of many community-developed or community-maintained software packages, including (in alphabetical order): Matplotlib \cite{Matplotlib}, NumPy \cite{NumPy}, and SciPy \cite{SciPy}. Bibliographic information was obtained from the NASA Astrophysical Data System.
\end{acknowledgments}

\appendix*
\section{}
In order to avoid the technical difficulties introduced by adiabatic regularization, we employ an alternative scheme which discards the vacuum contributions to the field energy-momentum tensor in their entirety. Albeit cruder, this method yields a good approximation to the field energy density $\rho(t)$ and pressure $P(t)$ in regimes dominated by particle production. Care must be taken, however, to ensure that the resulting expressions for these quantities satisfy the cosmological continuity equation. Below we demonstrate that this can be achieved by selecting an appropriate definition for the adiabatic vacuum.

The residual freedom that exists in the definition of the adiabatic mode functions allows for a slight shift in the balance between particle and vacuum contributions to energy density and pressure expressed in Eqs.~(\ref{Eq. Energy Density and Pressure}). Although small, this latitude can be exploited to ensure that the vacuum contributions $\rho_{\mathrm{vac}}(t)$ and $P_{\mathrm{vac}}(t)$ defined by Eqs.~(\ref{Eq. Energy and Pressure Vacuum}) independently satisfy the cosmological continuity equation.

Underlying this separation between particle and vacuum components are the functions $W_{k}(t)$ and $V_{k}(t)$. The first of these is given by the asymptotic series in Eq.~(\ref{Eq. Adiabatic W}), which is the solution to the differential equation
\begin{equation}\label{Eq. Differential Adiabatic W}
	 W_{k}^{2}(t) = \Omega_{k}^{2}(t) + \frac{3}{4} \frac{\dot{W}_{k}^{2}(t)}{W_{k}^{2}(t)} - \frac{1}{2} \frac{\ddot{W}_{k}(t)}{W_{k}(t)} \, ,
\end{equation}
obtained from the substitution of Eq.~(\ref{Eq. Adiabatic Modes}) into Eq.~(\ref{Eq. Mode Equation}). The function $V_{k}(t)$, on the other hand, encapsulates the remaining freedom in the definition of the adiabatic vacuum, and can be chosen to have any convenient functional form which satisfies the following constraint:
\begin{equation}\label{Eq. Condition Adiabatic V}
	 V_{k}(t) - H(t) < \mathcal{O}(k^{-2}) \quad \text{as} \quad k \rightarrow \infty \, .
\end{equation}
A natural choice which meets the above requirement is given by
\begin{equation}\label{Eq. Appendix Adiabatic V}
	V_{k}(t) = -\frac{\dot{W}_{k}(t)}{W_{k}(t)} \, .
\end{equation}

Interestingly, this functional form also guarantees that the vacuum contributions $\rho_{\mathrm{vac}}(t)$ and $P_{\mathrm{vac}}(t)$ satisfy the cosmological continuity equation:
it can be verified with the aid of Eqs.~(\ref{Eq. Differential Adiabatic W}) and (\ref{Eq. Appendix Adiabatic V}) that
\begin{equation}\label{Eq. Vacuum Cosmological Continuity}
	\dot{\rho}_{\mathrm{vac}}(t) = - 3H(t)\Big[ \rho_{\mathrm{vac}}(t) + P_{\mathrm{vac}}(t) \Big] \, ,
\end{equation}
where the right-hand side follows from the left-hand side by explicit calculation. It is worth noting that this result is valid for all truncation orders of $W_{k}(t)$ as given by Eq.~(\ref{Eq. Adiabatic W}).

Hence, provided the function $V_{k}(t)$ has the form established in Eq.~(\ref{Eq. Appendix Adiabatic V}), it follows directly from Eqs.~(\ref{Eq. Cosmological Continuity}) and (\ref{Eq. Vacuum Cosmological Continuity}) that the vacuum and particle contributions to the field energy density and pressure independently satisfy the cosmological continuity equation.

\bibliographystyle{apsrev4-1}
\bibliography{Backreaction1.bib}

\end{document}